\documentclass{article}
\usepackage{graphicx} 
\usepackage{amsmath,amsfonts}
\usepackage{algorithmic}
\usepackage{array}
\usepackage[caption=false,font=normalsize]{subfig}
\usepackage{textcomp}
\usepackage{stfloats}
\usepackage{url}
\usepackage{verbatim}
\usepackage{graphicx}
\usepackage{mathtools}
\usepackage{multirow}
\usepackage{balance}
\usepackage{cite}
\usepackage{booktabs}

\usepackage{array}
\usepackage{multirow}
\usepackage{longtable}
\usepackage{rotating}
\usepackage{enumerate}
\providecommand{\keywords}[1]
{
  \small	
  \textbf{Keywords:} #1
}
\date{}

\title{MCDAN: a Multi-scale Context-enhanced Dynamic Attention Network for Diffusion Prediction}
\author
{Xiaowen~Wang, 
Lanjun~Wang$^\ast$, 
Yuting~Su,
Yongdong~Zhang,
An-An~Liu$^\ast$
\thanks{Xiaowen~Wang, Lanjun~Wang, Yuting~Su, and An-An~Liu are with the Tianjin University, Tianjin 300072, China.
Xiaowen~Wang and An-An~Liu are also with the Institute of Artificial Intelligence, Hefei Comprehensive National Science Center, Hefei 230088, China. 
Yongdong~Zhang is with the University of Science and Technology of China, Hefei 230026, China.
(Corresponding authors: 
1. Lanjun~Wang, E-mail: wang.lanjun@outlook.com;
2. An-An~Liu, E-mail: anan0422@gmail.com).
	}
}

\begin{document}

\maketitle

\begin{abstract}
Information diffusion prediction aims at predicting the target users in the information diffusion path on social networks.
Prior works mainly focus on the observed structure or sequence of cascades, trying to predict to whom this cascade will be infected passively. In this study, we argue that user intent understanding is also a key part of information diffusion prediction. We thereby propose a novel Multi-scale Context-enhanced Dynamic Attention Network (MCDAN) to predict which user will most likely join the observed current cascades. Specifically, to consider the global interactive relationship among users, we take full advantage of user friendships and global cascading relationships, which are extracted from the social network and historical cascades, respectively. 
To refine the model's ability to understand the user's preference for the current cascade,
we propose a multi-scale sequential hypergraph attention module to capture the dynamic preference of users at different time scales. Moreover, we design a contextual attention enhancement module to strengthen the interaction of user representations within the current cascade. Finally, to engage the user’s own susceptibility, we construct a susceptibility label for each user based on user susceptibility analysis and use the rank of this label for auxiliary prediction. We conduct experiments over four widely used datasets and show that MCDAN significantly overperforms the state-of-the-art models. The average improvements are up to 10.61\% in terms of  Hits@100 and 9.71\% in terms of MAP@100, respectively.
\end{abstract}


\section{Introduction}\label{introduction}
Information diffusion prediction, which is also known as cascade prediction, is a challenging but critical task in many real-world application domains, such as influence maximization~\cite{kumar2022influence},  influential user detection~\cite{oro2017detecting}, online advertising~\cite{wang2020social}, recommendation~\cite{huang2015social,zhao2017social,sang2020context,feng2022relation} and rumor detection~\cite{zeng2022early}. Recent works on diffusion prediction rely on the achievements of deep neural networks, jointly learning the structure of the social graph and dynamic diffusion graph through Graph Neural Networks (GNN)~\cite{yang2019multi, yuan2021dyhgcn, sun2022ms}.

\begin{figure}[h]
  \centering
  \includegraphics[width=\linewidth]{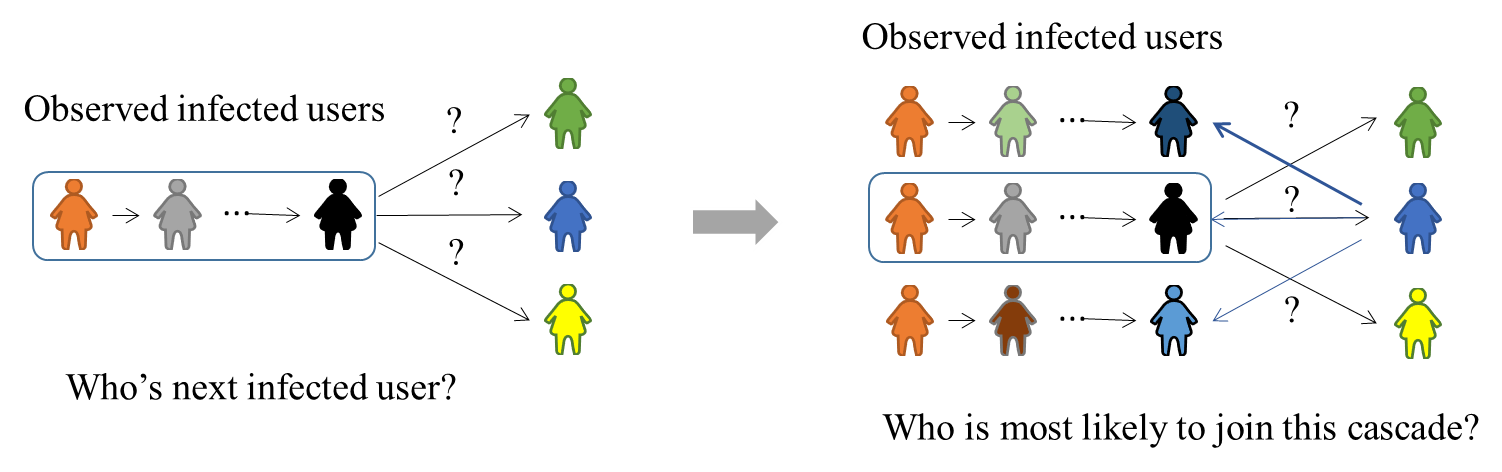}
  \caption{Illustrative examples for traditional next infected user prediction (left) and user-aware next infected user prediction (right).}
  \label{fig:fig1}
\end{figure}

However, as shown in Fig.~\ref{fig:fig1}(left), existing works always focus on predicting to whom this cascade will be infected~\cite{yang2019multi}. In this context, there is no difference between information diffusion and virus transmission, where users are always passively infected. This is not consistent with the fact that information diffusion is a two-way selection process. 
Fig.~\ref{fig:fig1}(right) shows a realistic scenario on social networks, in which a user receives a lot of information every day, but he/she only selects a few of them to forward and/or to reply.   That is to say, 
from the user's perspective, when the cascade appears within the user's visible range, the user makes a decision on whether to join or not based on their own intention for this cascade. As a result, instead of predicting who is the next infected user, the task is more appropriately expressed as ‘who is most likely to join this cascade’. Apart from unidirectional information infection, we argue that user intent understanding is also a key part of information diffusion prediction.

The challenges of understanding user intention in information diffusion prediction lie in accurately portraying the target user portrait, which consists of user interaction, cascade memory, and user susceptibility. 
First, user interaction refers to interactive relationships between the target user and others. Since information diffusion is built on social networks, most of the existing methods take advantage of the social connections~\cite{yang2019multi, yuan2021dyhgcn, sun2022ms}. Although the who-follow-who relations can represent friendship, it is not enough or exact to represent global dependencies. 
Second, cascade memory refers to the representation of the current cascade. Previous works capture users' dynamic preference through time series slicing and look up the memory for each cascade~\cite{yuan2021dyhgcn, sun2022ms}. 
Regrettably, these models suffer from information loss due to their reliance on a single time scale.
Meanwhile, the context interaction within the cascade also needs to be captured, which cannot be learned only by looking up the stored memory.
Third, user susceptibility refers to the personal attribute of the target user regarding whether he/she is susceptible to infection. 
Although some works consider the social role~\cite{yang2015rain} or social influence~\cite{li2013modeling}, they are not straightforward to describe user susceptibility.  To the best of our knowledge, no previous works consider the concept of user susceptibility. 

To address these challenges, we propose a Multi-scale Context-enhanced Dynamic Attention Network (MCDAN) to achieve the user-aware next infected user prediction. 
First, to more accurately represent the global dependencies of users, we do not only take advantage of their friendships from the social network but also consider global cascading relationships, which come from historical cascades. The intention behind this is that people are more likely to interact with friends as well as people who have interacted with them before.
Second, to refine the model's ability to understand the user's preferences for the current cascade,
we propose a multi-scale sequential hypergraph attention module to capture the dynamic preference of users at different time scales and design a contextual attention enhancement module to strengthen the interaction of user representations within the current cascade. 
Third, to present the user’s own susceptibility, we construct a susceptibility label for each user based on user susceptibility analysis and use the rank of this label for the final auxiliary prediction.

In summary, the main contributions are as follows:
\begin{itemize}
    \item We propose a user-aware framework named MCDAN which aims at predicting the target user in the information diffusion path by understanding the user’s intention. Through this framework, we fully utilize user portraits composed of user interaction, cascade memory, and user susceptibility.

    \item We build a global cascading graph based on historical cascades, which patches the friendship graph and learns the global user interaction. 
    
    \item We propose a multi-scale sequential hypergraph attention module to encode the multi-scale cascade memory. 
    
    \item To enhance the context dependency of the current cascade, a contextual attention enhancement module is also proposed after memory look-up. 
    
    \item We assign susceptibility labels to users based on historical cascades. The labels ranking helps lock the target user.

    \item Extensive experiments demonstrate the superiority of our method. We outperform state-of-the-art baseline methods on four public datasets with average improvements of 10.61\% in  Hits@100 and 9.71\% in  MAP@100, respectively. In addition, we conduct ablation studies to demonstrate the effectiveness of each part and parameter analysis experiments to discuss the sensitivity of different key parameters. 
\end{itemize}

The rest of this paper is organized as follows. In Section~\ref{related work}, we briefly review the related works including traditional models and deep learning models for information diffusion prediction. In Section~\ref{method}, we define the problem and introduce the proposed MCDAN model. In Section~\ref{experiments}, we report all the results of comparative experiments, ablation study experiments, and parameter analysis experiments. Finally, we summarize the paper in Section~\ref{conclusion}.

\section{Related Work}\label{related work}
\subsection{Information Diffusion Prediction}
Information diffusion prediction is to predict the trajectories as well as the participants in information spreading in the future based on observed cascades and relevant known information. So far, there are many methods for modeling and predicting information cascading and data types related to cascading~\cite{liao2019popularity, gou2018learning, jia2018predicting, yang2019neural, islam2018deepdiffuse, wang2018sequential, sankar2020inf, yuan2021dyhgcn, sun2022ms}. Here we categorize them into two categories: traditional models and deep learning models.

Traditionally, earlier works mainly focus on feature-based models~\cite{liao2019popularity, gou2018learning, jia2018predicting} and generative models~\cite{wang2017linking, samanta2017lmpp, zhao2020online}. 
In feature-based models, different features can be extracted from given information through feature engineering, and classical machine learning methods are used for prediction~\cite{lai2020hyfea,wang2020feature}. Most features are manually constructed, focusing on extracting features from information content~\cite{chen2022and,wu2022deeply}. 
In generative models, the spreading of information is widely characterized by probabilistic statistical generative approaches such as epidemic models~\cite{zhao2020online} and stochastic point processes~\cite{wang2017linking, samanta2017lmpp}. 
However, feature-based models are not generalizable to different scenarios~\cite{zhou2021survey}, which are inefficient in large-scale networks. Although generative models are easily applicable to modeling information diffusion, they mainly help process modeling but are less powerful in making accurate predictions.

With advances in deep neural networks, prior works utilize or expand related networks to learn potential information~\cite{yang2019neural, islam2018deepdiffuse, wang2018sequential, sankar2020inf, yuan2021dyhgcn, sun2022ms},  which can be classified into cascades diffusion based models and social graph based models. 
For the cascades diffusion based models, typically, DeepDiffuse combines the LSTM network and attention mechanism to learn time information~\cite{islam2018deepdiffuse}. SNIDSA incorporates a structure attention module and gating mechanism into a recurrent neural network (RNN) for integrating the structural and sequential information~\cite{wang2018sequential}.
For the social graph based models, they are with an intuition behind that people have common interests with their friends. FOREST introduces social relationships through GNN~\cite{yang2019multi}. Inf-VAE embeds social homophily into the prediction model~\cite{sankar2020inf}. With a deeper understanding of information diffusion, DyHGCN jointly learns the structure of the social graph and dynamic diffusion graph~\cite{yuan2021dyhgcn}. MS-HGAT further introduces diffusion hypergraphs into user representation learning, among which designs a sequential hypergraph attention network to learn user preference dynamically~\cite{sun2022ms}. However, existing works neglect the complete portrayal of user portraits, resulting in information loss in the global dependency of users, the integrity of information, and the user's own susceptibility.

\subsection{User Portrait Modeling}
User portrait is a concept first proposed by Alan Cooper, and it is interpreted as a concrete representation of the target user~\cite{cooper1999inmates}.
The modeling of user portraits aims to reflect user intention through data analysis, including behaviour patterns and interest preference~\cite {li2022modeling}.

Researches on user portrait mainly focus on three directions~\cite{chen2021multi}, which are user attribute annotation~\cite{zeng2016user}, user preference understanding~\cite{xing2016user,wu2016modeling,gao2016context}, and user behavior analysis~\cite{li2019machine}.
For user attribute annotation, existing works collect some feature information through social annotation systems~\cite {zeng2016user}. User portrait is then built on the extracted features~\cite{mueller2016gender,gu2018modeling}.
For user preference understanding, existing models include user preference prediction~\cite{wu2016modeling} and similar user mining~\cite{ma2012habit,huang2015social}. The user preference prediction model builds user portraits based on historical information in dynamic social networks~\cite{wu2016modeling}. The similar user mining model searches for similar users by mining the same user habits from mobile devices~\cite{ma2012habit}.
For user behaviour analysis, the existing methods learn from historical behaviours, establish user behaviour profiles, and analyze potential relationships among users~\cite{li2010behaviour,zhao2016user,gu2018modeling}. The methods are mainly applied in fields such as marketing~\cite{saura2021using} and recommendation~\cite{sun2019bert4rec}.

In our study, we are the first to apply the concept of user portrait to information diffusion prediction. The purpose of user portrait modeling is to understand user intention in information diffusion scenarios, with key elements being user susceptibility, information portrayal, and the relationships between users.

\section{Method}\label{method}

\begin{figure*}[ht]
  \centering
  \includegraphics[width=\textwidth]{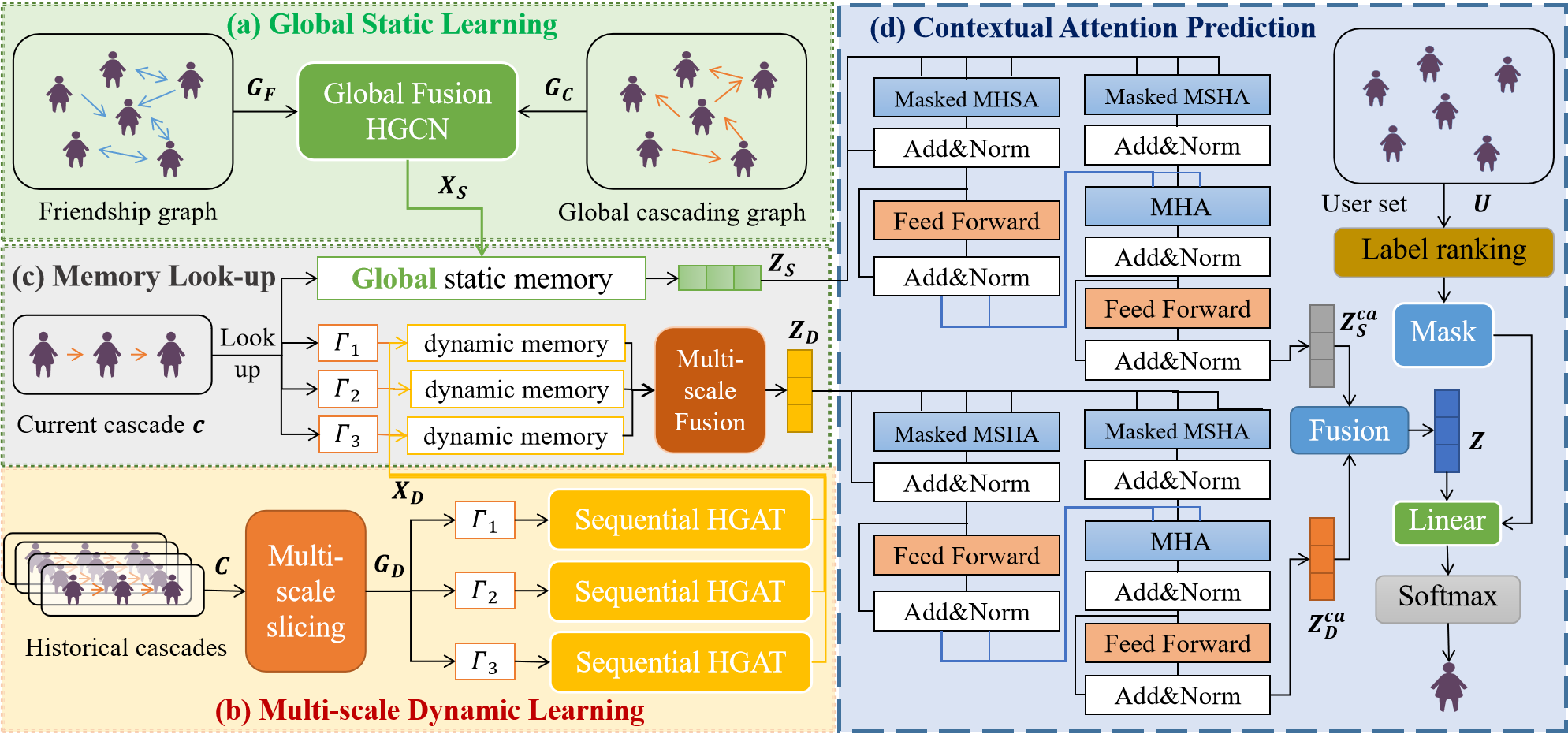}
  \caption{An overview of the proposed MCDAN framework. It comprises four components: (a) Global Static Learning, which aims at learning the global users' dependencies;  (b) Multi-scale Dynamic Learning, which aims at storing the dynamic users' preference memory based on hypergraphs over $M$ different time scales(we set $M = 3$ for example in Fig.~\ref{fig:fig2}); (c) Memory Look-up, which aims at depicting the representations for the observed current cascade and (d) Contextual Attention Prediction, which aims at encoding the cascade with a CAE module and predicting the next user based on user susceptibility analysis.}
  \label{fig:fig2}
\end{figure*}

\subsection{Problem Formulation}\label{problem_formulation}
Since information diffusion prediction aims at predicting the future diffusion process based on the current cascades and relevant knowledge~\cite{sun2022ms}, the task is defined as follows.
Suppose that a collection of diffusion cascades $C$ is propagated among a set of users $U$. The user set is recorded as $U$ = $\{u_1, u_2, ..., u_N\}$, where $N$ denotes the maximum user number. The collection of diffusion cascades is recorded as $C = \{c_1, c_2, c_3,..., c_H\}$, where $H$ denotes the maximum historical cascade number. 
Given a friendship graph $G_F$ = ($U$, $E_F$) where $E_F$ means the friendship edges and an observed current cascade $c=\{(u_i,t_i) \vert u_i \in U\}$ where $t_i$ means the time $u_i$ join the current cascade. The target is to estimate the probability $\hat{y}$ of each user from the $U$ joining this cascade $c$ at the next step and predict the candidate by ranking all the probabilities.

\subsection{Framework}
In this study, we apply the social network as a friendship graph $G_F$ and historical cascades $C$ to construct a global cascading graph $G_C$. Meanwhile, we construct diffusion hypergraphs $G_D$ which are based on different time scales.

The overall framework of the proposed MCDAN is depicted in Fig.~\ref{fig:fig2}. As shown, the model consists of four modules, namely Global Static Learning, Multi-scale Dynamic Learning, Memory Look-up and Contextual Attention Prediction.
\begin{enumerate}[(a)]
\item Global Static Learning module aims at learning the user interaction and storing the global static memory $X_S$ from the friendship graph $G_F$ and the global cascading graph $G_C$. 
\item Multi-scale Dynamic Learning module aims at constructing multi-scale hypergraphs $G_D$ and storing the dynamic users' preference memory $X_D$ based on the hypergraphs.
\item Memory look-up module is used to depict the representations for the current cascade from static memory $X_S$ and dynamic memory $X_D$, which learned from Global Static Learning module and Multi-scale Dynamic Learning module, respectively. The output of this module is two embeddings named global static user embedding $Z_S$ and multi-scale dynamic user embedding $Z_D$.
\item Contextual Attention Prediction module then encodes global static user embedding $Z_S$ and multi-scale dynamic user embedding $Z_D$ with a Contextual Attention Enhancement(CAE) module and applies an insusceptibility mask to obtain the final output. 
\end{enumerate} 

We then introduce each module in detail in the following sections.

\subsection{Global Static Learning}\label{gsl}

\begin{figure}[h]
  \centering
  \includegraphics[width=0.5\linewidth]{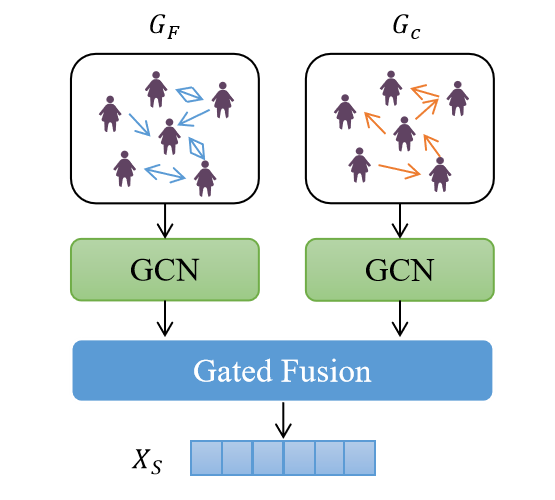}
  \caption{Illustration of the proposed Global Fusion HGCN. The friendship graph and global cascading graph are fed into GCN to obtain user friendship representation and user cascading representation, respectively. Then, the representations are fused to obtain the final global static user representation.}
  \label{fig:fig3}
\end{figure}

Since the social network represents the friendship and cascading graph from historical cascades represents the historical interaction, we learn the global users' dependencies from the related prior knowledge. 
Specifically, the global cascading graph is recorded as $G_C$ = ($U$, $E_C$), where $E_C$ means the historical cascading edges.
Given the friendship graph $G_F$ and global cascading graph $G_C$, we feed them into the Global Fusion Heterogeneous Graph Convolutional Networks(HGCN) as shown in Fig.~\ref{fig:fig3}. 

Specifically, we use two multi-layer graph convolutional networks (GCN)~\cite{kipf2016semi} to learn user friendship representation and user cascading representation from the two graphs, respectively. 
The layer-wise propagation rule can be defined as follows:
\begin{equation}\label{eq1}
X_F{(l+1)}=ReLU(\widetilde{D}_F^{\--{\frac{1}{2}}}\widetilde{A}_F\widetilde{D}_F^{\--{\frac{1}{2}}}{X_F}{(l)}{W_F})
\end{equation}
where $X_F{(0)} \in \mathbb{R}^{N \times d}$ is randomly initialized user friendship embedding with normal distribution, $\widetilde{A}_F = A_F + I$ is built based on the adjacency matrix $A_F$ of the given graph $G_F$, $\widetilde{D}_F$ is the corresponding degree matrix, and $W_F$ is a trainable weight matrix. 


We can obtain the user cascading representation $X_C$ through GCN in a similar layer-wise propagation rule as follows:
\begin{equation}\label{eq2}
X_C{(l+1)}=ReLU(\widetilde{D}_C^{\--{\frac{1}{2}}}\widetilde{A}_C\widetilde{D}_C^{\--{\frac{1}{2}}}{X_C}{(l)}{W_C})
\end{equation}
where $X_C{(0)} \in \mathbb{R}^{N \times d}$ is randomly initialized user cascading embedding with normal distribution, $ \widetilde{A}_C = A_C + I$ is built based on the adjacency matrix $A_C$ of $G_C$, $ \widetilde{D}_C$ denotes degree matrix, and $W_C$ is a trainable weight matrix. 

Finally, we fuse user friendship representation $X_F$ and user cascading representation $X_C$ to obtain the final global static memory $X_S$. The gated fusion module is depicted as follows:
\begin{equation}\label{eq3}
X_S = \alpha X_F + (1\--{\alpha})X_C
\end{equation}
\begin{equation}\label{eq4}
\alpha = \frac{exp(W_S^T\sigma(W_1X_F))}{exp(W_S^T\sigma(W_1X_F))+exp(W_S^T\sigma(W_1X_C))}
\end{equation}
where $X_S \in \mathbb{R}^{N \times d}$, $\sigma(\cdot)$ represents the $tanh$ activation function, $W_1$ denotes the transformation matrix and $W_S$ is the vector of attention which both $W_1$ and $W_S$ are trainable. 

\subsection{Multi-scale Dynamic Learning}\label{mdl}
Although the static memory describes the users' dependencies globally, it is not enough to reflect the temporal user interaction relationships. 
As shown in Fig.~\ref{fig:fig2}(b), we further construct hypergraphs based on all the historical cascades over different time scales. Then, we apply the Sequential Hypergraph Attention Network(HGAT) module to learn dynamic user interactions from the hypergraphs.

\subsubsection{Multi-scale slicing}
We construct the diffusion hypergraph set $G_D$ based on historical cascades $C$ over $M$ different time scales.
In each time scale, we arrange the cascades in chronological order and split the historical diffusion timeline into $\Gamma$ time intervals. 
The multi-scale hypergraphs are constructed as follows:
\begin{equation}\label{eq5}
 \begin{aligned}
&G_D=\{G_D^\Gamma \vert \Gamma=\Gamma_1,\Gamma_2,...,\Gamma_M\} \\
&G_D^\Gamma=\{G_D^\tau = (U^\tau, E_D^\tau)\vert \tau=1,2,3,..., \Gamma\}
\end{aligned}
\end{equation}
where $M$ denotes the number of time scales, $\Gamma$ denotes number of time intervals {which can be set as $\{\Gamma_1,\Gamma_2,...,\Gamma_M\}$ and the total timeline is cut into $\Gamma$ pieces of time intervals,} $G_D^\tau$ denotes the diffusion hypergraph of the {$\tau$-th} time interval, $U^\tau$ and $E_D^\tau$ denote the corresponding users and hyperedges of the users, respectively. 

Here we go through this process with a brief example. Suppose that $M$ is 2 and corresponding $\Gamma$ are \{4, 8\}, we have 100 cascades with 8 minutes.
Firstly, for $\Gamma=4$, we divide the cascades into 4 subsets with 2 minutes each in chronological order. Here the length of the time interval is 2 minutes. Note that since there are 100 cascades in total, each subset contains 100 subsequences at most.
Secondly, we construct the hypergraph based on the subsequences of each time interval. In each hypergraph, once the users of the same cascade appear in the same time interval, a hyperedge is built. Also note that since there are 100 subsequences at most, the maximum number of hyperedges in one hypergraph is 100.
Thirdly, we repeat the above steps with $\Gamma = 8$, in which the number of subsets is 8 and the length of the time interval is 1 minute. 
Finally, we obtain a single-scale set of 4 hypergraphs and a single-scale set of 8 hypergraphs corresponding to $M=2$ and $\Gamma = \{4, 8\}$.



\subsubsection{Sequential HGAT}
{Given  the diffusion hypergraph set $G_D$ with a size $M$,}
we apply $M$ Sequential HGAT modules to learn $M$ single-scale dynamic user memories. 
The final output of the modules is depicted as follows:
 \begin{equation}\label{eq6}
 \begin{aligned}
 & X_D=\{X_D^{\Gamma}\vert \Gamma =\Gamma_1,\Gamma_2,...,\Gamma_M\} \\
 & X_D^\Gamma = Sequential~HGAT(G_D^\Gamma)
 \end{aligned}
\end{equation}
where $X_D^\Gamma$ denotes the single-scale dynamic memory learned from a single-scale set of diffusion hypergraphs  $G_D^\Gamma$ through a Sequential HGAT module.

Specifically, a Sequential HGAT module consists of $\Gamma$ blocks. Each block corresponds to a hypergraph for each time interval.
\begin{figure}[h]
  \centering
  \includegraphics[width=0.8\linewidth]{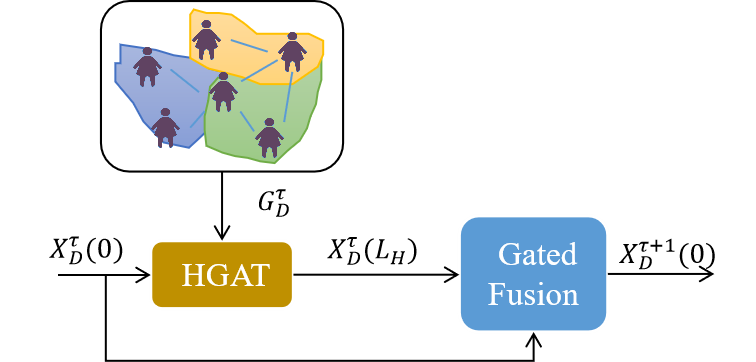}
  \caption{Illustration of a Sequential HGAT block. In each block, an $L_{H}$-layer HGAT module and a gated fusion module are applied.}
  \label{fig:fig4}
\end{figure}
In each block, as shown in Fig.~\ref{fig:fig4}, we first apply an $L_{H}$-layer HGAT to model the high-order interaction among users from a hypergraph for each time interval, 
the process can be formulated as follows:
\begin{equation}\label{eq7}
X_D^\tau(L_{H})=HGAT(X_D^\tau(0),G_D^\tau)
\quad \tau = 1, 2, 3,.., \Gamma
\end{equation}
where $G_D^\tau$ denotes the corresponding diffusion hypergraph in the time interval $\tau$, $X_D^\tau(0)$ denotes the initial user representation, and the output $X_D^\tau(L_{H})$ is stored in single-scale memory $X_D^\Gamma$ described in Equation(\ref{eq6}).

Then, as shown in Fig.~\ref{fig:fig4}, we connect $X_D^\tau(0)$ and $X_D^\tau(L_{H})$ with a gated fusion module in chronological order to obtain the initial user representation of the next time interval as follows:
\begin{equation}\label{eq10}
X_D^{\tau+1}(0)=g_fX_D^\tau(0)+(1\--{g_f})X_D^\tau(L_{H}) 
\end{equation}
\begin{equation}\label{eq11}
g_f = \frac{exp(W_{DH}^T\sigma(W_gX_D^\tau(0)))}{exp(W_{DH}^T\sigma(W_gX_D^\tau(0)))+exp(W_{DH}^T\sigma(W_gX_D^\tau(L_{H})))}
\end{equation}
where $W_g$ denotes the transformation matrix and $W_{DH}$ is the vector of attention. Note that we use the friendship representation $X_F$ stored in global static memory as the initial user representation of HGAT in the first time interval to warm up the Sequential HGAT module, i.e. $X_D^{1}(0)=X_F$.

\begin{figure}[h]
  \centering
  \includegraphics[width=0.7\linewidth]{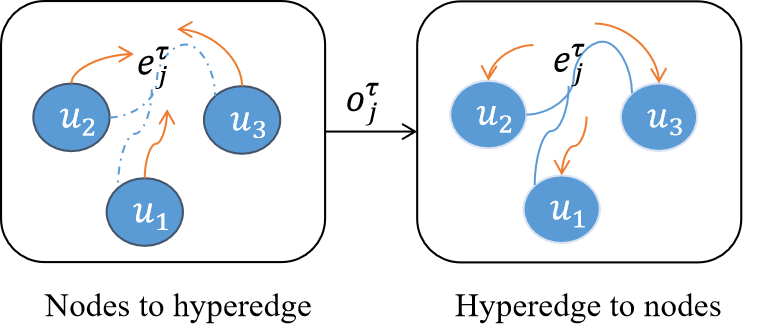}
  \caption{Illustration of an $L_{H}$-layer HGAT block in the hyperedge $e_j^\tau$ of $G_D^\tau$.}
  \label{fig:fig5}
\end{figure}

Furthermore, we illustrate the process of HGAT in detail. Given the diffusion hypergraph $G_D^\tau$,
in each $L_{H}$-layer HGAT, we have two steps to obtain the high-order interaction among users. 
That is, (i) nodes to single hyperedge and (ii) hyperedges to nodes. Fig.~\ref{fig:fig5} shows the learning process of a single hyperedge $e_j^\tau$  in the diffusion hypergraph $G_D^\tau$.

\begin{enumerate}[(i)]
\item  Nodes to single hyperedge:
The first step of HGAT aims to learn the representation $o_j^\tau$ of hyperedge $e_j^\tau$ by aggregating the initial user representation $x_i^\tau$ of all the connected nodes $u_i^\tau$:
\begin{equation}\label{eq8}
o_j^\tau(l+1)= ReLU(\sum_{u_i^\tau \in e_j^\tau} W_{h_1}x_i^\tau(l))
\end{equation}
where $o_j^\tau(l+1)$ denotes the learned representation of the hyperedge $e_j^\tau$, $W_{h_1}$ denotes the trainable parameter, and $x_i^\tau(l)$ denotes the initial user representation.

\item   Hyperedges to nodes:
The second step is based on the learned representations of hyperedges, we integrate all the hyperedges $E_{D, i}^\tau$ joined by $u_i^\tau$ in the time interval $\tau$ to update the user representation $x_i^\tau$:
\begin{equation}\label{eq9}
x_i^\tau(l+1)= ReLU(\sum_{e_j^\tau \in E_{D,i}^\tau} W_{h_2}o_j^\tau(l+1))
\end{equation}
where $x_i^\tau(l+1)$ denotes the updated user representation, and $W_{h_2}$ denotes the trainable parameter. 
Note that all the final updated user representations i.e. $x_i^\tau(L_{H})$ constitute the output $X_D^\tau(L_{H})$ described in Equation(\ref{eq7}).
\end{enumerate}

\subsection{Memory Look-up}\label{memorylookup}
In this section, we depict the current cascade $c$ with the global static memory $X_S$ obtained from Equation(\ref{eq3}) and multi-scale dynamic memory $X_D$ obtained from Equation(\ref{eq6}). 

As shown in Fig.~\ref{fig:fig2}(c),
for the static memory, we arrange it with the order of user in the given cascade $c=\{(u_i,t_i) \vert u_i \in U\}$, thus obtaining global static user embedding:
$Z_S=\{x_S^i\vert u_i\in U, x_S^i\in X_S\}$, where $Z_S\in  \mathbb{R}^{|c| \times d}$.

For the dynamic memory, we first obtain each single-scale dynamic user embedding and then fuse them with a multi-scale fusion module.

To obtain the single-scale dynamic user embedding from each single-scale dynamic memory $X_D^\Gamma$, we query the corresponding users’ representation at the nearest time interval based on the given cascade $c=\{(u_i,t_i) \vert u_i \in U\}$ as well. Since $t_i$ denotes the time that $u_i$ join and if it is divided into the time interval $\tau$, the representation can be represented as:
$Z_D^\Gamma=\{x_D^{i,\tau}\vert u_i\in U, t_i\in \tau,  x_D^{i,\tau}\in X_D^\Gamma\}$, where $Z_D^\Gamma\in \mathbb{R}^{|c| \times d}$.

After obtaining $M$ single-scale dynamic user embeddings, we propose a multi-scale fusion module to obtain the final multi-scale dynamic user embedding $Z_D \in \mathbb{R}^{|c| \times d}$:
\begin{equation}\label{eq12}
\begin{aligned}
&Z_D= \sum_{\Gamma \in \{\Gamma_1,\Gamma_2,...,\Gamma_M\}} {m_\Gamma Z_D^\Gamma} \\
&m_\Gamma = \frac{exp(W_{D}^T\sigma(W_mZ_D^{\Gamma}))}{\sum_{\Gamma}{exp(W_{D}^T\sigma(W_mZ_D^{\Gamma}))}}
\end{aligned}
\end{equation}
where $W_m$ denotes the transformation matrix and $W_{D}$ is the vector of attention. 

\subsection{Contextual Attention Prediction}\label{cap}
In order to further capture the context information within the current cascade, as shown in Fig.~\ref{fig:fig2}(d), we propose a Contextual Attention Enhancement(CAE) module to enhance the context dependency. Based on the user susceptibility analysis, we then feed the context-enhanced embeddings to obtain the final predicted user. We illustrate the process in detail in the following sections.

\subsubsection{CAE module}
Instead of using an RNN~\cite{yang2019multi} or a single masked Multi-Head Self-Attention(MHSA) module~\cite{yuan2021dyhgcn, sun2022ms} to decode the given user embeddings, we design an encode-decode module to learn the context information.

\begin{enumerate}[(i)]
\item  Self-attention encode:
Given the static user embedding $Z_S$, we first apply a masked MHSA module to learn the hidden embeddings $h_S$:
\begin{equation}\label{eq14}
    \begin{aligned}
& MaskedAtt(Q, K, V) = softmax(\frac{QK^T}{\sqrt{d_\Omega}}+\mathbb{M})V,\\
& h_\omega = MaskedAtt(Z_SW_\omega^Q, Z_SW_\omega^K, Z_SW_\omega^V),\\
& h_S = [h_1; h_2;...; h_\Omega] W^O \\
    \end{aligned}
\end{equation}
where $W_\omega^Q$, $W_\omega^K$, $W_\omega^V$ and $W^O$ are trainable parameters, $d_\Omega = d/\Omega$, $d$ is the dimension of the embedding and $\Omega$ denotes the number of heads of attention. The mask matrix $\mathbb{M}$ is defined as $\mathbb{M}_{ij}=0$ if $i \le j$ otherwise $-\infty$, which is used to avoid label leakage. 

Then, we obtain the attentive representation as context embedding $Z_S^c$ through the two layers fully-connected neural network named Feed Forward network:
\begin{equation}\label{eq15}
Z_S^c= ReLU(h_S W_{E_1}+b_1) W_{E_2}+b_2
\end{equation}
where $W_{E_1}$ and $W_{E_2}$ are trainable matrices, $b_1$ and $b_2$ are trainable bias parameters.
Note that residual connection is realized by the Add \& Norm layer.

\item Context-attention decode:
The context-attention decoder has three steps:

First, we apply a masked MHSA module to obtain the masked hidden embedding $h_S^c$,  $h_S^c = h_S$ where $h_S$ is from Equation(\ref{eq14}). 

Second, we feed the masked hidden embedding $h_S^c$ and context embedding $Z_S^c$ into a multi-head attention(MHA) module to obtain the context-enhanced hidden embedding $h_S^{ca}$. Specifically, we set $h_S^c$ as $Q$, $Z_S^c$ as the $K$ and $V$.
The process can be formulated as follows:
\begin{equation}\label{eq16}
    \begin{aligned}
& Att(Q, K, V) = softmax(\frac{QK^T}{\sqrt{d_\Omega}})V,\\
& h_\omega^{ca} = Att(h_S^cW_\omega^{Q^{ca}}, Z_S^cW_\omega^{K^{ca}}, Z_S^cW_\omega^{V^{ca}}),\\
& h_S^{ca} = [h_1^{ca}; h_2^{ca};...; h_\Omega^{ca}] W^{O^{ca}} \\
    \end{aligned}
\end{equation}
where $W_\omega^{Q^{ca}}$, $W_\omega^{K^{ca}}$, $W_\omega^{V^{ca}}$ and $W^{O^{ca}}$ are trainable matrices.

Third, we apply a Feed Forward network to learn the context-enhanced user embedding $Z_S^{ca}$:
\begin{equation}\label{eq17}
Z_S^{ca}= ReLU(h_S^{ca} W_{E_3}+b_3) W_{E_4}+b_4
\end{equation}
where $W_{E_3}$ and $W_{E_4}$ are trainable matrices, $b_3$ and $b_4$ are trainable bias parameters.
\end{enumerate}

Meanwhile, given the dynamic user embedding $Z_D$, we obtain the dynamic context-enhanced user embedding $Z_D^{ca}$ in a similar way.

After then, we incorporate the static context-enhanced user embedding $Z_S^{ca}$ and the dynamic context-enhanced user embedding $Z_D^{ca}$ to obtain the final context-enhanced user embedding $Z$. We implement a new gated fusion module to realize: 
\begin{equation}\label{eq18}
Z = \beta Z_S^{ca} + (1\--{\beta})Z_D^{ca}
\end{equation}
\begin{equation}\label{eq19}
\beta = \frac{exp(W_Z^T\sigma(W_2Z_S^{ca}))}{exp(W_Z^T\sigma(W_2Z_S^{ca}))+exp(W_Z^T\sigma(W_2Z_D^{ca}))}
\end{equation}
where $W_2$ denotes the transformation matrix and $W_Z$ is the vector of attention.

\subsubsection{Prediction based on user susceptibility}
Before the final prediction, we construct a susceptibility label for each user based on all the historical cascades. Specifically, we calculate the frequency of infection for each user based on the given cascades. 
Then we match the user susceptibility score based on this frequency, thus obtaining the susceptibility label.

Based on the susceptibility label, we sort the user set $U$ with the label ranking. We set an insusceptible threshold to put users who rank behind it as insusceptible users.

Finally, we compute the diffusion probabilities with the mask of the insusceptible users:
\begin{equation}\label{eq20}
\hat{y}=softmax(W_pZ+Mask)
\end{equation}
where $W_p$ denotes the trainable parameter, and $Mask$ is used to mask users who are set as insusceptible.

During the training, we apply the cross-entropy loss as the objective function:
\begin{equation}\label{eq21}
 Loss(\theta) = -\sum_{t=2}^{|c|}\sum_{i=1}^{N}y_{ti}log(\hat{y}_{ti})
\end{equation}
where $\theta$ denotes all the parameters to be trainable. If the user $u_i$ join in cascade $c$ at the step $t$, $y_{ti} = 1$, otherwise $y_{ti} = 0$.

\section{Experiments}\label{experiments}
In this section, we first illustrate the datasets, implementation details and baseline models used in our experiments, and then we report the results of comparative experiments, ablation study experiments, and parameter analysis experiments to demonstrate the effectiveness of our proposed MCDAN model.

\subsection{Experiment Setting}
\subsubsection{Datasets}
Following the previous work~\cite{sun2022ms}, we conduct our experiments on four publicly available datasets, i.e., Twitter~\cite{hodas2014simple}, Douban~\cite{zhong2012comsoc}, Android~\cite{sankar2020inf} and Christianity~\cite{sankar2020inf}. The first two datasets are collected from social media, the last two are collected from Stack-Exchanges. 

For a fair comparison, we preprocess the four datasets according to the method in~\cite{sun2022ms} by removing cascades with lengths beyond 200. The statistics of the preprocessed datasets are shown in Table~\ref{tab:dataset}. Specifically, $\# Fri. Links$  denotes the number of edges in the friendship graph from the social network. 
$\# Cas. Links$ denotes the number of edges in the global cascading graph which is built based on the historical cascades. 
Note that all isolated edges are not counted.

For each dataset, we randomly split the dataset by 8:1:1 for training, validation, and testing.

\begin{table}
  \caption{Statistics of the preprocessed datasets in our experiments}
  \label{tab:dataset}
   \resizebox{\linewidth}{!}{
  \begin{tabular}{lrrrr}
    \toprule
    Datasets&	Twitter&	Douban	& Android	&Christianity\\
    \midrule
   \# Users &	12,627 & 12,232 &	2,927 &	1,651\\
   \# Fri. Links &	309,631 &	198,496 &	24,459	& 21,955\\
   \# Cas. Links &	73,036	&51,797	&23,958	&11,328\\
   \# Cascades	& 3,442	& 3,475&	678 &	589\\
   Avg. Length	& 32.60	& 21.76	& 42.05	& 26.02\\
  \bottomrule
\end{tabular}
    }
\end{table}

\subsubsection{Implementation Details}\label{setting_detail}
We implement the proposed MCDAN via PyTorch~\cite{Ketkar2021}. Specifically, we apply Adam as the optimizer, and the learning rate is initialized as 0.01. The batch size of training and  the dimension of embeddings are both set to 64. 

For global static learning (Sec.~\ref{gsl}), we apply two two-layer GCNs to learn the friendship graph and global cascading graph, respectively. For multi-scale dynamic learning(Sec.~\ref{mdl}), the number of time scales $M$ is set to 3 and the corresponding numbers of time intervals are set to \{4, 8, 16\}. During each time scale, one-layer HGAT is adopted to learn the high-order interaction from each hypergraph. For contextual attention prediction(Sec.~\ref{cap}), the number of heads in multi-head attention $\Omega$ is set to 14. The insusceptible label threshold $ratio_t$ is set to adapt to the datasets. Here we set it to 12\% for Twitter, 6.9\% for Douban, 0.17\% for Android and 0.15\% for Christianity. 

\subsubsection{Evaluation Metrics}
Following previous works~\cite{yang2019multi,sankar2020inf,yuan2021dyhgcn,sun2022ms}, the evaluation metrics we use in this study are two ranking metrics, that is, Hits score on top k (Hits@k) and Mean Average Precision on top k(MAP@k), where k = \{10, 50, 100\}.

\subsection{Baselines}
To verify the effectiveness of the proposed MCDAN, we compare it with the following baselines in two categories.
\subsubsection{Cascades Diffusion based}
These methods infer future diffusion processes based on the given cascades without social information.
\begin{itemize}
\item DeepDiffuse~\cite{islam2018deepdiffuse} combines the LSTM network and attention mechanism to model the diffusion path.

\item Topo-LSTM~\cite{wang2017topological}  extends standard LSTM to Topo-LSTM to model the cascade diffusion.

\item NDM~\cite{yang2019neural} models the cascades by self-attention mechanism and CNNs.

\item SNIDSA~\cite{wang2018sequential} integrates structural attention modules and gating mechanisms into RNN for model learning.
\end{itemize}

\subsubsection{Social Graph based}
These methods utilize social information and given cascades for diffusion prediction.
\begin{itemize}
\item FOREST~\cite{yang2019multi} incorporates the social connections through graph neural networks into RNN for prediction.

\item Inf-VAE~\cite{sankar2020inf} learns social homophily by utilizing graph neural network architectures and integrates it into a variational autoencoder.

\item DyHGCN~\cite{yuan2021dyhgcn} jointly learns the graph representations of the social graph and diffusion graph for dynamic diffusion modeling.

\item MS-HGAT~\cite{sun2022ms} proposes a memory-enhanced sequential hypergraph attention network on the basis of the social graph and diffusion hypergraphs. 
\end{itemize}

\subsection{Overall performance}\label{overall}

\begin{table*}[ht]
 \caption{Overall results with Hits@k scores for k = 10, 50, 100 on four public datasets(\%). The average improvement is up to 10.61\% in terms of the Hits@100 score.}
  \label{tab:hits_all}
 \resizebox{\linewidth}{!}{
\begin{tabular}{ccccccccccccc}
\toprule
\multirow{2}{*}{model} & \multicolumn{3}{c}{Twitter}                      & \multicolumn{3}{c}{Douban}                       & \multicolumn{3}{c}{Android}                      & \multicolumn{3}{c}{Christianity}                 \\
 \cline{2-4}
 \cline{5-7}
 \cline{8-10}
 \cline{11-13}
                       & @10            & @50            & @100           & @10            & @50            & @100           & @10            & @50            & @100           & @10            & @50            & @100           \\
\midrule
DeepDiffuse & 5.79 &10.80 &18.39& 9.02 &14.93& 19.13& 4.13 &10.58 &17.21 &10.27 &21.83 &30.74\\
Topo-LSTM & 8.45 & 15.80 & 25.42 & 8.57 & 16.53 & 21.47 & 4.56 & 12.63 & 16.53 & 12.28 & 22.63 & 31.52\\
NDM & 15.21 & 28.23 & 32.30 &10.00 & 21.13 & 30.14 & 4.85 & 14.24 & 18.97&15.41& 31.36 &45.86\\
SNIDSA & 25.37 & 36.64 & 42.89 & 16.23 & 27.24 & 35.59 &5.63 &15.22& 20.93 &17.74 &34.58 &48.76\\
\midrule
FOREST & 28.67 & 42.07 & 49.75 & 19.50 & 32.03 & 39.08 &9.68 &17.73 &24.08 &24.85 &42.01& 51.28\\
Inf-VAE & 14.85 & 32.72 & 45.72 & 8.94 & 22.02 & 35.72 & 5.98 & 14.70 & 20.91 & 18.38 & 38.50 & 51.05\\
DyHGCN & 31.88 & 45.05 & 52.19 & 18.71 & 32.33 & 39.71 & 9.10 & 16.38 & 23.09 & 26.62 & 42.80 &52.47\\
MS-HGAT                & 33.50          & 49.59          & 58.91          & 21.33          & 35.25          & 42.75          & 10.41          & 20.31          & 27.55          & 28.80          & 47.14          & 55.62          \\
\midrule
\textbf{MCDAN(ours)}           & \textbf{38.45} & \textbf{55.78} & \textbf{64.25} & \textbf{49.39} & \textbf{58.58} & \textbf{62.81} & \textbf{11.89} & \textbf{25.10} & \textbf{32.79} & \textbf{35.49} & \textbf{56.92} & \textbf{67.41}\\
\bottomrule
\end{tabular}
    }
\end{table*}

\begin{table*}[]
 \caption{Overall results with MAP@k scores for k = 10, 50, 100 on four public datasets(\%). The average improvement is 9.71\% in terms of the MAP@100 score.}
  \label{tab:map_all}
 \resizebox{\linewidth}{!}{
\begin{tabular}{ccccccccccccc}
\toprule
\multirow{2}{*}{model} & \multicolumn{3}{c}{Twitter}                      & \multicolumn{3}{c}{Douban}                       & \multicolumn{3}{c}{Android}                      & \multicolumn{3}{c}{Christianity}                 \\
 \cline{2-4}
 \cline{5-7}
 \cline{8-10}
 \cline{11-13}
& @10 & @50  & @100   & @10     & @50      & @100  & @10   & @50     & @100    & @10    & @50       & @100           \\
\midrule
DeepDiffuse & 5.87 & 6.80 & 6.39 & 6.02 & 6.93 & 7.13 & 2.30 & 2.53 & 2.56 & 7.27 & 7.83 & 7.84\\
Topo-LSTM & 8.51 & 12.68 & 13.68 & 6.57 & 7.53 & 7.78 & 3.60 & 4.05 & 4.06 & 7.93 & 8.67 & 9.86\\
NDM & 12.41 & 13.23 & 14.30 & 8.24 & 8.73 & 9.14 & 2.01 & 2.22 & 2.93 & 7.41 & 7.68 & 7.86\\
SNIDSA & 15.34 & 16.64 & 16.89 & 10.02 & 11.24 & 11.59 & 2.98 & 3.24 & 3.97 & 8.69 & 8.94 & 9.72\\
\midrule
FOREST & 19.60 & 20.21 & 21.75 & 11.26 & 11.84 & 11.94 & 5.83 & 6.17 & 6.26 & 14.64 & 15.45 & 15.58\\
Inf-VAE & 19.80 & 20.66 & 21.32 & 11.02 & 11.28 & 12.28 & 4.82 & 4.86 & 5.27 & 9.25 & 11.96 & 12.45\\
DyHGCN & 20.87 & 21.48 & 21.58 & 10.61 & 11.26 & 11.36 & 6.09 & 6.40 & 6.50 & 15.64 & 16.30 & 16.44\\
MS-HGAT  & 22.49   & 23.17        & 23.30       & 11.72     & 12.52      & 12.60    & 6.39     & 6.87     & 6.96         & 17.44     & 18.27       & 18.40        \\
\midrule
\textbf{MCDAN(ours)}                     & \textbf{25.89}     & \textbf{26.69}     & \textbf{26.81}     & \textbf{40.70}     & \textbf{41.13}     & \textbf{41.19}    & \textbf{7.47}      & \textbf{8.04}      & \textbf{8.15}      & \textbf{22.88}       & \textbf{23.78}       & \textbf{23.94}     
\\
\bottomrule
\end{tabular}
}
\end{table*}

The comparison results over four datasets are reported in Table~\ref{tab:hits_all} and Table~\ref{tab:map_all}, which correspond to two metrics, Hits@k and MAP@k, respectively.
Since MS-HGAT~\cite{sun2022ms} is the SOTA model, all the experimental results of baselines reported are cited from it.
From these two tables, we obtain the following observations:

\begin{enumerate}[1.]
\item Our MCDAN model achieves optimal predictive performance. Compared with the SOTA model MS-HGAT~\cite{sun2022ms}, we have achieved better performance on all four datasets with average improvements of 10.61\% in the Hits@100 score and 9.71\% in the MAP@100 score, respectively.
Especially, for the Douban dataset, our MCDAN reaches up to 20.06\% improvement in Hits@100 score and 28.59\% in MAP@100 score.

\item  Regarding the Hits@100 metric shown in Table~\ref{tab:hits_all}, our method improves 5.34\% in Twitter, 20.06\% in Douban, 5.24\% in Android and 11.79\% in Christianity by comparing with the SOTA model, respectively. Since the Hits@100 metric refers to the hitting rate of the first 100 results, the reason for such improvement is our method presents more global interaction and context information at different scales, which helps learn more potential interactions from both historical cascades and the current cascade. 

\item Regarding the MAP@100 metric shown in Table~\ref{tab:map_all}, our method improves 3.51\% in Twitter, 28.59\% in Douban, 1.19\% in Android and 5.54\% in Christianity by comparing with the SOTA model, respectively. Since the MAP@100 metric refers to the mean average precision of the first 100 results, the reason for such improvement is our method of user susceptibility analysis as an auxiliary prediction can eliminate more erroneous candidates and improve the precision of retrieval. In addition, the enrichment of the potential interaction also improves prediction precision.
\end{enumerate}

\subsection{Ablation study}
We conduct ablation studies on the different components of the proposed MCDAN model over the four public datasets. The variants of the model are designed as:

\textbf{w/o  G}lobal cascading graph:  remove global cascading graph. The Equation(\ref{eq3}) is replaced by:
\begin{equation}\label{eq22}
X_S = X_F
\end{equation}

\textbf{w/o  M}ulti-scale diffusion hypergraphs: replace the multi-scale diffusion hypergraphs with single-scale diffusion hypergraphs. Here we set the number of time intervals to 8, which is the same as the previous models~\cite{yuan2021dyhgcn,sun2022ms}. The Equation(\ref{eq12}) is replaced by:
 \begin{equation}\label{eq23}
Z_D=Z_D^\Gamma 
\end{equation}
where $\Gamma$ = 8.

\textbf{w/o  C}AE module:  replace the contextual attention enhancement module with only a multi-head self-attention decoder for each user embedding. The Equation(\ref{eq18}) is replaced by:
\begin{equation}\label{eq24}
Z= \gamma Z_S^{c} + (1\--{\gamma})Z_D^{c}
\end{equation}
\begin{equation}\label{eq25}
\gamma = \frac{exp(W_\gamma^T\sigma(W_3Z_S^{c}))}{exp(W_\gamma^T\sigma(W_3Z_S^{c}))+exp(W_\gamma^T\sigma(W_3Z_D^{c}))}
\end{equation}
where $Z_S^{c}$ denotes the user static embedding obtained through Equation(\ref{eq15}), $Z_D^{c}$ denotes the user dynamic embedding obtained in a similar way from $Z_D$, $W_3$ denotes the transformation matrix and $W_\gamma$ is the vector of attention.

\textbf{w/o  L}abel ranking mask: remove the final insusceptible label mask. The Equation(\ref{eq20}) is replaced by:
\begin{equation}\label{eq26}
\hat{y}=softmax(W_pZ)
\end{equation}

\begin{table*}[]
 \caption{Ablation study with Hits@k scores for k = 10, 50, 100 on four public datasets(\%). Note that we use underlining to mark the results of the most effective component.}
  \label{tab:hits_ab}
 \resizebox{\linewidth}{!}{
\begin{tabular}{ccccccccccccc}
\toprule
\multirow{2}{*}{model} & \multicolumn{3}{c}{Twitter}                      & \multicolumn{3}{c}{Douban}                       & \multicolumn{3}{c}{Android}                      & \multicolumn{3}{c}{Christianity}                 \\
 \cline{2-4}
 \cline{5-7}
 \cline{8-10}
 \cline{11-13}
& @10 & @50  & @100   & @10     & @50      & @100  & @10   & @50     & @100    & @10    & @50       & @100           \\

\midrule
\textbf{MCDAN}     & \textbf{38.45} & \textbf{55.78} & \textbf{64.25} & \textbf{49.39} & \textbf{58.58} & \textbf{62.81} & \textbf{11.89} & \textbf{25.10} & \textbf{32.79} & \textbf{35.49} & \textbf{56.92} & \textbf{67.41}\\
\midrule
w/o  G	&\underline{32.02} &\underline{49.94}	&\underline{60.27}	&28.97	&42.72	&49.11 &\underline{10.88}	&22.61	&29.99	&32.59	&\underline{51.12} &64.96\\
w/o  M	&38.16	&54.96	&63.17	&39.65	&52.43	&57.58 &11.58	&22.92	&30.61	&\underline{31.47}	&52.46	&62.05\\
w/o  C	&35.41	&50.49	&58.55	&\underline{20.50}	&\underline{34.86}	&\underline{41.97} &11.11	&21.45	&28.67	&31.92	&52.46	&\underline{61.38}\\
w/o  L	&34.94	&51.64	&60.58	&45.77	&55.64	&60.05 &11.34	&\underline{21.06}	&\underline{28.44}	&32.14	&52.90	&66.29\\
\bottomrule
\end{tabular}
}
\end{table*}

\begin{table*}[ht]
 \caption{Ablation study with MAP@k scores for k = 10, 50, 100 on four public datasets(\%). Note that we use underlining to mark the results of the most effective component.}
  \label{tab:map_ab}
 \resizebox{\linewidth}{!}{
\begin{tabular}{ccccccccccccc}
\toprule
\multirow{2}{*}{model} & \multicolumn{3}{c}{Twitter}                      & \multicolumn{3}{c}{Douban}                       & \multicolumn{3}{c}{Android}                      & \multicolumn{3}{c}{Christianity}                 \\
 \cline{2-4}
 \cline{5-7}
 \cline{8-10}
 \cline{11-13}
& @10 & @50  & @100   & @10     & @50      & @100  & @10   & @50     & @100    & @10    & @50       & @100           \\

\midrule
\textbf{MCDAN} &\textbf{25.89}     & \textbf{26.69}     & \textbf{26.81}     & \textbf{40.70}     & \textbf{41.13}     & \textbf{41.19}    & \textbf{7.47}      & \textbf{8.04}      & \textbf{8.15}      & \textbf{22.88}       & \textbf{23.78}       & \textbf{23.94}    
\\
\midrule
w/o  G	&\underline{19.90} &\underline{20.72}	&\underline{20.87}	&19.68	&20.30	&20.39 &\underline{6.63}	&\underline{7.17}	&\underline{7.27}	&20.34	&21.23 &21.42\\
w/o  M	&25.18	&25.96	&26.08	&29.89	&30.50	&30.57 &6.99	&7.50	&7.61	&20.20	&21.14	&21.28\\
w/o  C	&23.59	&24.29	&24.40	&\underline{11.17}	&\underline{11.87}	&\underline{11.97}  &6.85	&7.29	&7.39	&\underline{19.26}	&\underline{20.16}	&\underline{20.28}\\
w/o  L	&22.73	&23.49	&23.62	&37.36	&37.81	&37.87  &7.09	&7.52	&7.62	&19.64	&20.52	&20.71\\
\bottomrule
\end{tabular}}
\end{table*}

The ablation results are reported in Table~\ref{tab:hits_ab} and Table~\ref{tab:map_ab} which are related to two metrics, Hits@k and MAP@k. From these two tables, we obtain the following observations:

\begin{enumerate}[1.]
\item Our MCDAN model achieves the best performance over all the datasets, which confirms that all components of our model are effective.

\item Regarding social media-based datasets i.e. Twitter and Douban, the requirements for different components in different datasets are not completely consistent. As the Hits@k metric shown in Table~\ref{tab:hits_ab} and the MAP@k metric shown in Table~\ref{tab:map_ab}, we observe that for the Twitter dataset, the most effective component is the global cascading graph, the least useful one is the multi-scale diffusion hypergraphs. 
However, for the Douban dataset, the most effective component is the CAE module, and the least powerful one is the masking of insusceptibility labels. The difference shows the different propagation tendencies of different datasets. Specifically, the Twitter dataset exhibits global propagation consistency, while the Douban dataset focuses more on the contextual information of the current sequence.

\item Regarding Stack-Exchanges-based datasets i.e. Android and Christianity, the conclusion is consistent with results on the social media-based datasets. Specifically, we observe that for the Android dataset, the least powerful component is the multi-scale diffusion hypergraphs. While for the Christianity dataset, the least effective one is the masking of insusceptibility labels. However, compared with social media-based datasets, the performance differences of variant models on these two datasets are not significant, which may be limited by the size of the datasets.
\end{enumerate}

\begin{figure*}[ht]
\centering  
\subfloat[Hits@k on Twitter]{  
\begin{minipage}{6cm}
\centering    
\includegraphics[scale=0.5]{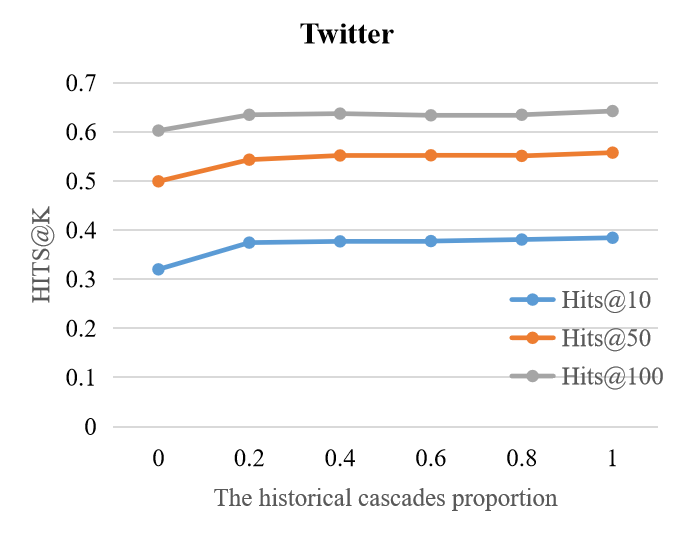} 
\end{minipage}
}
\subfloat[Hits@k on Douban]{  
\begin{minipage}{6cm}
\centering    
\includegraphics[scale=0.5]{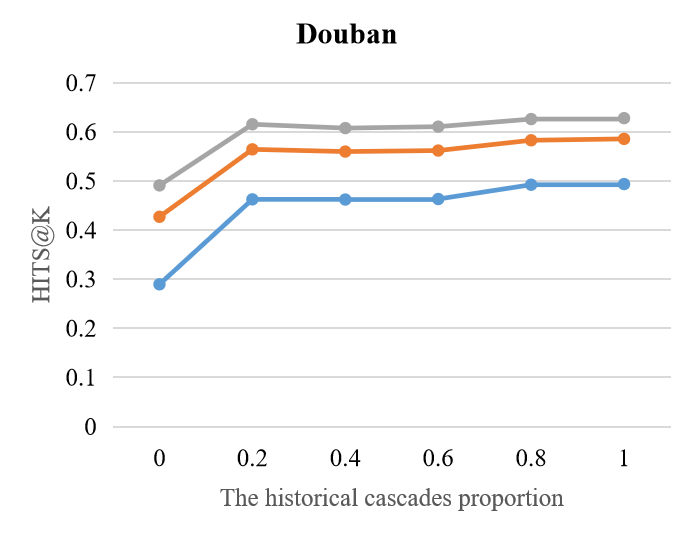} 
\end{minipage}
}
\\
\subfloat[Hits@k on Android]{  
\begin{minipage}{6cm}
\centering    
\includegraphics[scale=0.5]{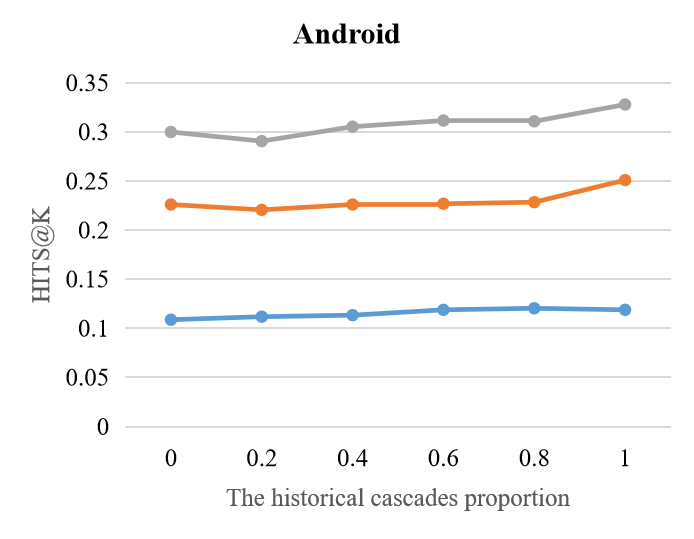} 
\end{minipage}
}
\subfloat[Hits@k on Christianity]{  
\begin{minipage}{6cm}
\centering    
\includegraphics[scale=0.5]{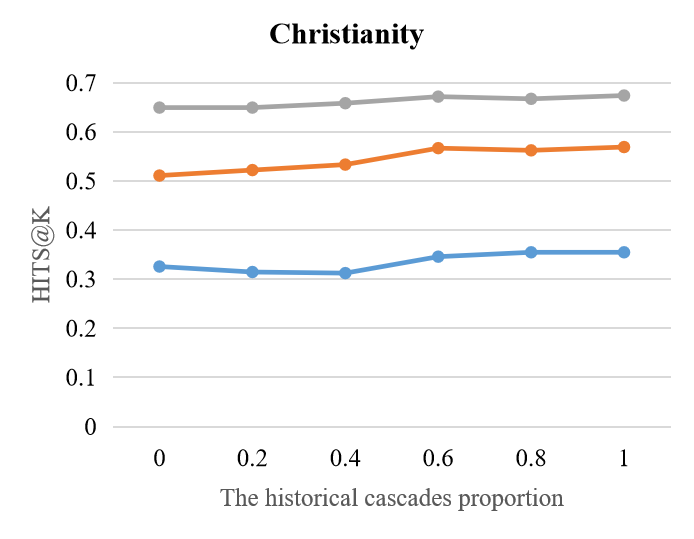} 
\end{minipage}
}
\\
\subfloat[MAP@k on Twitter]{
\begin{minipage}{6cm}
\centering    
\includegraphics[scale=0.5]{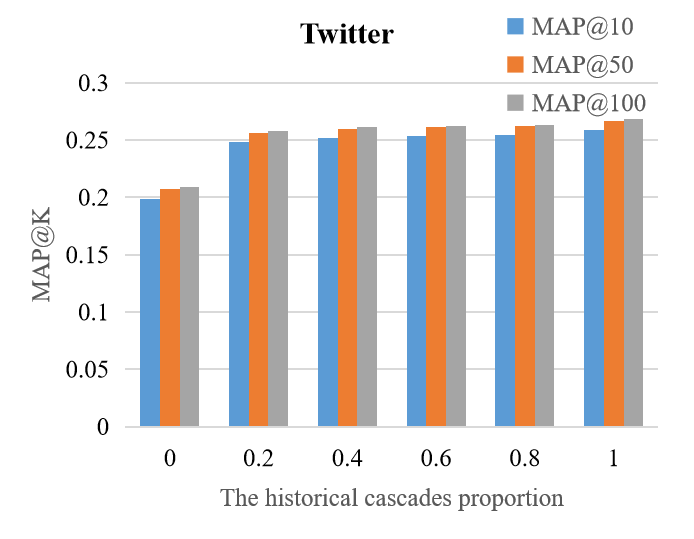}
\end{minipage}
}
\subfloat[MAP@k on Douban]{
\begin{minipage}{6cm}
\centering    
\includegraphics[scale=0.5]{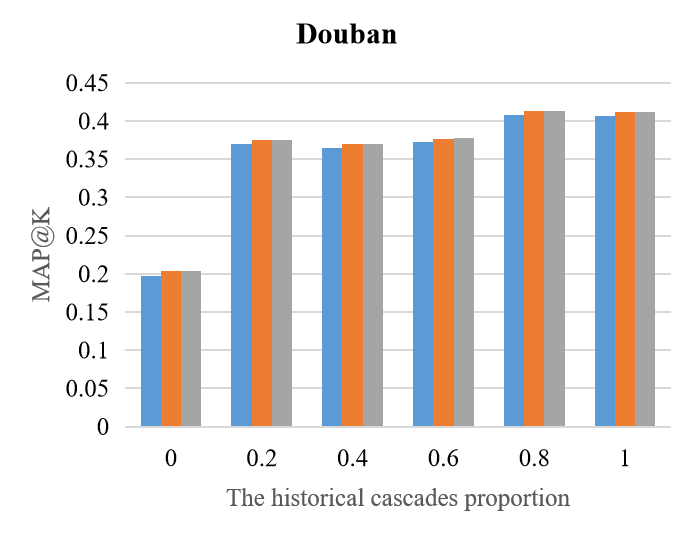}
\end{minipage}
}
\\
\subfloat[MAP@k on Android]{
\begin{minipage}{6cm}
\centering    
\includegraphics[scale=0.5]{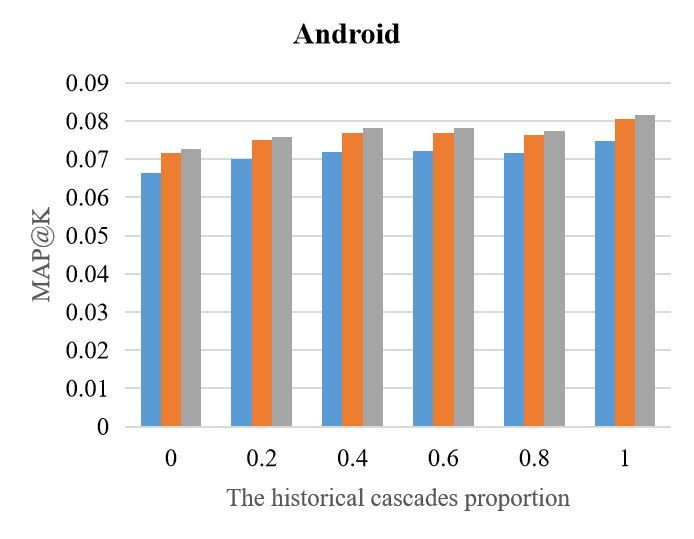}
\end{minipage}
}
\subfloat[MAP@k on Christianity]{
\begin{minipage}{6cm}
\centering    
\includegraphics[scale=0.5]{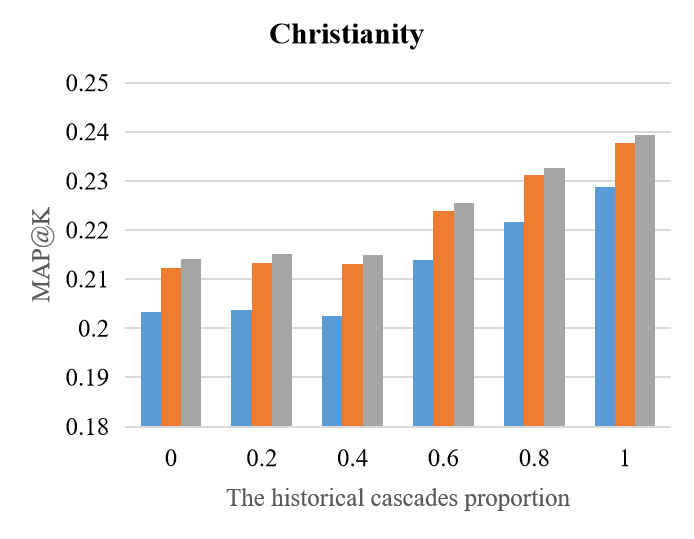}
\end{minipage}
}
\caption{Results of the impact of the historical cascades proportion on the four public datasets.}   

\label{fig:fig6}    
\end{figure*}

\begin{table*}[]
 \caption{Results of the impact of the number of time scales $M$ on four public datasets(\%). (Hits@k scores for k = 10, 50, 100)}
  \label{tab:hits_pa_num}
 \resizebox{\linewidth}{!}{
\begin{tabular}{cccccccccccccc}
\toprule
 \multirow{2}{*}{M}& \multirow{2}{*}{$\Gamma$} & \multicolumn{3}{c}{Twitter}                      & \multicolumn{3}{c}{Douban}                       & \multicolumn{3}{c}{Android}                      & \multicolumn{3}{c}{Christianity}                 \\
 \cline{3-5}
 \cline{6-8}
 \cline{8-11}
 \cline{12-14}
& & @10 & @50  & @100   & @10     & @50      & @100  & @10   & @50     & @100    & @10    & @50       & @100           \\

\midrule
1 & 8	& 38.16&	54.96	&63.17&	39.65&	52.43	&57.58&	11.58&	22.92&	30.61&	31.47	&52.46	&62.05\\
\midrule
\textbf{3} & \textbf{4,8,16}    & \textbf{38.45} & \textbf{55.78} & \textbf{64.25} & \textbf{49.39} & \textbf{58.58} & \textbf{62.81} & \textbf{11.89} & \textbf{25.10} & \textbf{32.79} & \textbf{35.49} & \textbf{56.92} & \textbf{67.41}\\
\midrule
5 &2,4,8,16,32	&34.86	&52.04	&61.25	&45.11	&54.96	&59.88	&10.65	&22.92	&31.24	&32.37	&50.89	&65.63
\\
\bottomrule
\end{tabular}
}
\end{table*}

\begin{table*}[]
 \caption{Results of the impact of the number of time scales $M$ on four public datasets(\%). (MAP@k scores for k = 10, 50, 100)}
  \label{tab:map_pa_num}
 \resizebox{\linewidth}{!}{
\begin{tabular}{cccccccccccccc}
\toprule
\multirow{2}{*}{M} & \multirow{2}{*}{$\Gamma$} & \multicolumn{3}{c}{Twitter}                      & \multicolumn{3}{c}{Douban}                       & \multicolumn{3}{c}{Android}                      & \multicolumn{3}{c}{Christianity}                 \\
\cline{3-5}
 \cline{6-8}
 \cline{8-11}
 \cline{12-14}
& & @10 & @50  & @100   & @10     & @50      & @100  & @10   & @50     & @100    & @10    & @50       & @100           \\

\midrule
1	& 8 &25.18	&25.96	&26.08	&29.89	&30.50	&30.57 &6.99	&7.50	&7.61	&20.20	&21.14	&21.28\\
\midrule
\textbf{3} & \textbf{4,8,16}&\textbf{25.89}     & \textbf{26.69}     & \textbf{26.81}     & \textbf{40.70}     & \textbf{41.13}     & \textbf{41.19}    & \textbf{7.47}      & \textbf{8.04}      & \textbf{8.15}      & \textbf{22.88}       & \textbf{23.78}       & \textbf{23.94}    
\\
\midrule
5	& 2,4,8,16,32 &22.90&	23.68	&23.82	&34.79	&35.26	&35.33	&6.88	&7.44	&7.56	&20.27	&21.22	&21.43\\
\bottomrule
\end{tabular}}
\end{table*}

\begin{table*}[]
 \caption{insusceptible label threshold setting.}
  \label{tab:set_lr}
 \resizebox{\linewidth}{!}{
\begin{tabular}{ccccccccccccccccc}
\toprule
 \multirow{2}{*}{$t_{ratio}$}& \multicolumn{4}{c}{Twitter}                      & \multicolumn{4}{c}{Douban}                       & \multicolumn{4}{c}{Android}                      & \multicolumn{4}{c}{Christianity} 
\\

 \cline{2-5}
 \cline{6-9}
 \cline{10-13}
 \cline{14-17}

& 0.02 & 0.04  & 0.06  & 0.08   & 0.02 & 0.04  & 0.06  & 0.08 & 0.02 & 0.04  & 0.06  & 0.08 & 0.02 & 0.04  & 0.06  & 0.08   \\

\midrule

\# insusceptible users&	252	&505&	757	&1010	&244	&489	&734	&978	&58&	117	&175	&234	&33	&66	&99	&132
\\
\bottomrule
\end{tabular}
}
\end{table*}

\begin{figure*}[htbp]
\centering  
\subfloat[Hits@k on Twitter]{  
\begin{minipage}{6cm}
\centering    
\includegraphics[scale=0.5]{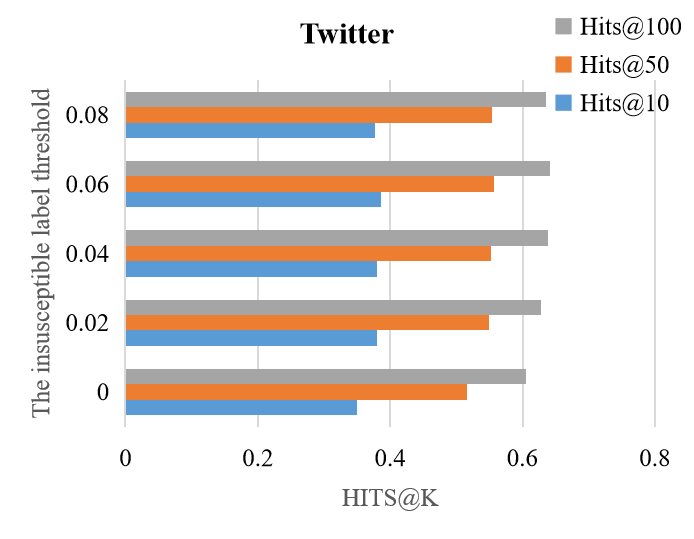} 
\end{minipage}
}
\subfloat[Hits@k on Douban]{  
\begin{minipage}{6cm}
\centering    
\includegraphics[scale=0.5]{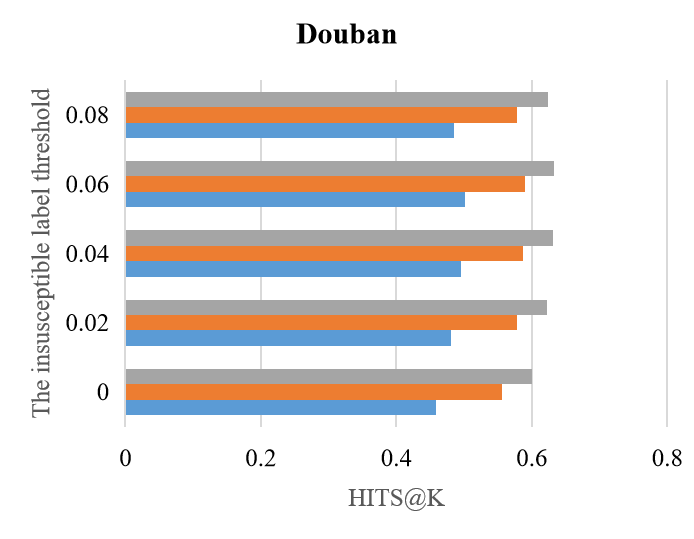} 
\end{minipage}
}
\\
\subfloat[Hits@k on Android]{  
\begin{minipage}{6cm}
\centering    
\includegraphics[scale=0.5]{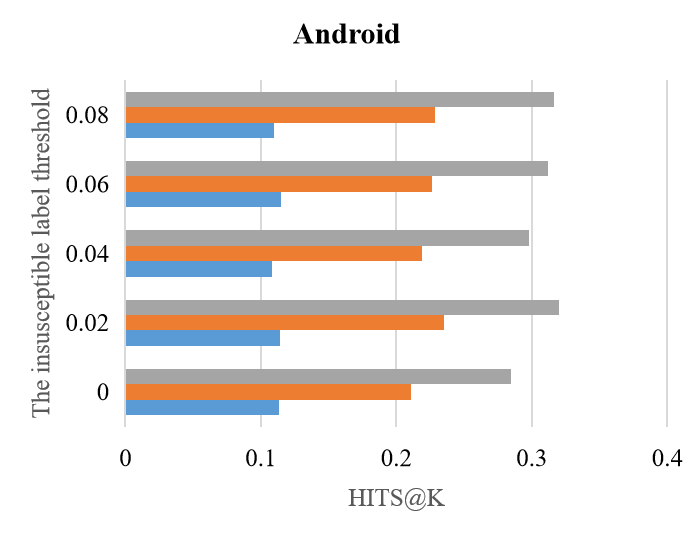} 
\end{minipage}
}
\subfloat[Hits@k on Christianity]{  
\begin{minipage}{6cm}
\centering    
\includegraphics[scale=0.5]{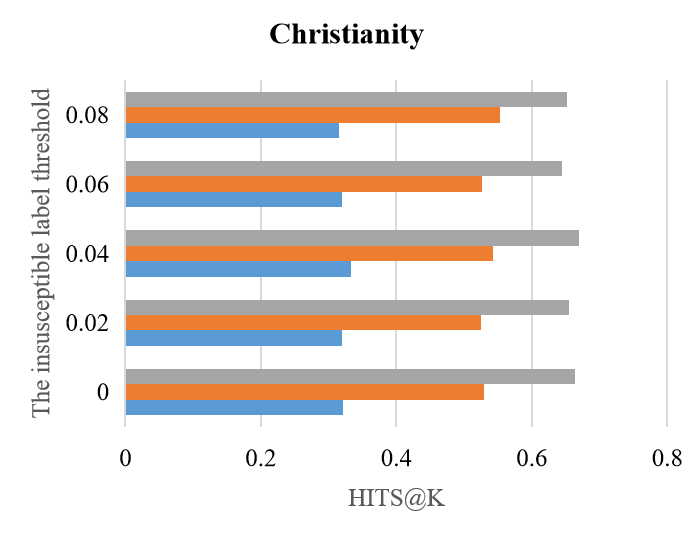} 
\end{minipage}}
\\
\subfloat[MAP@k on Twitter]{
\begin{minipage}{6cm}
\centering    
\includegraphics[scale=0.5]{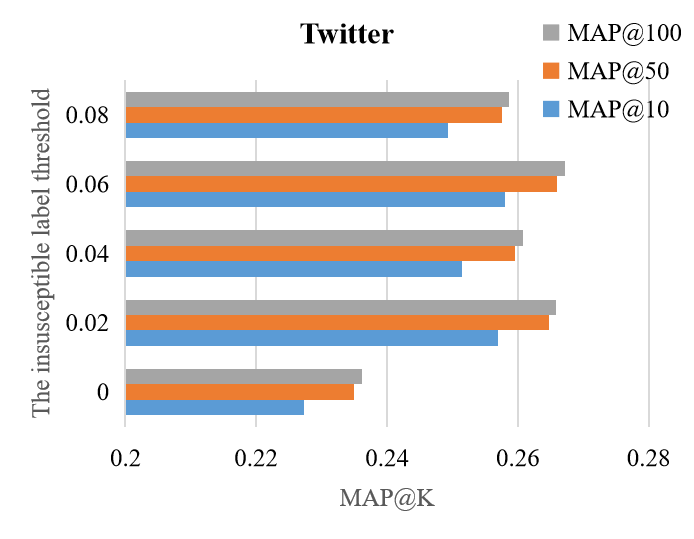}
\end{minipage}
}
\subfloat[MAP@k on Douban]{
\begin{minipage}{6cm}
\centering    
\includegraphics[scale=0.5]{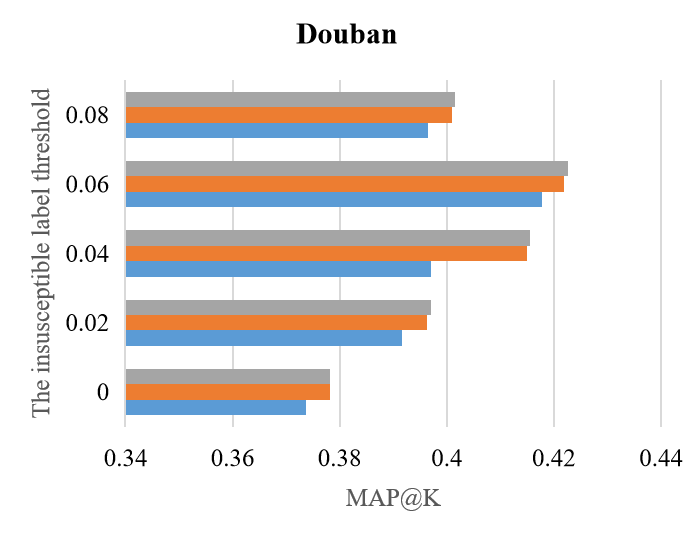}
\end{minipage}
}
\\
\subfloat[MAP@k on Android]{
\begin{minipage}{6cm}
\centering    
\includegraphics[scale=0.5]{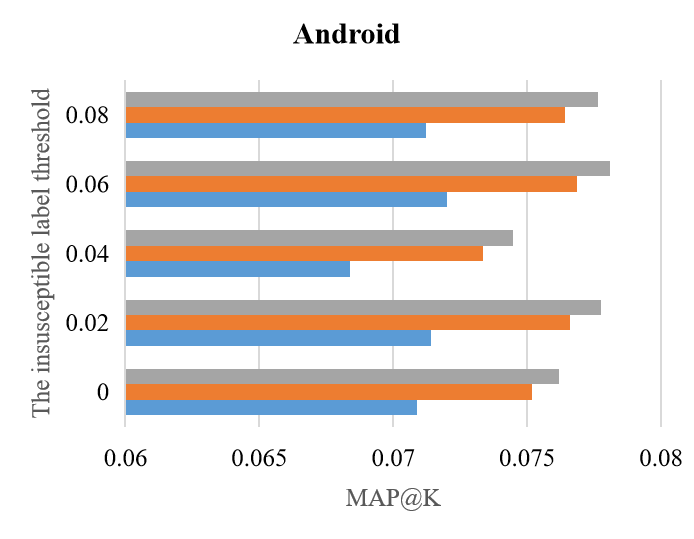}
\end{minipage}
}
\subfloat[MAP@k on Christianity]{
\begin{minipage}{6cm}
\centering    
\includegraphics[scale=0.5]{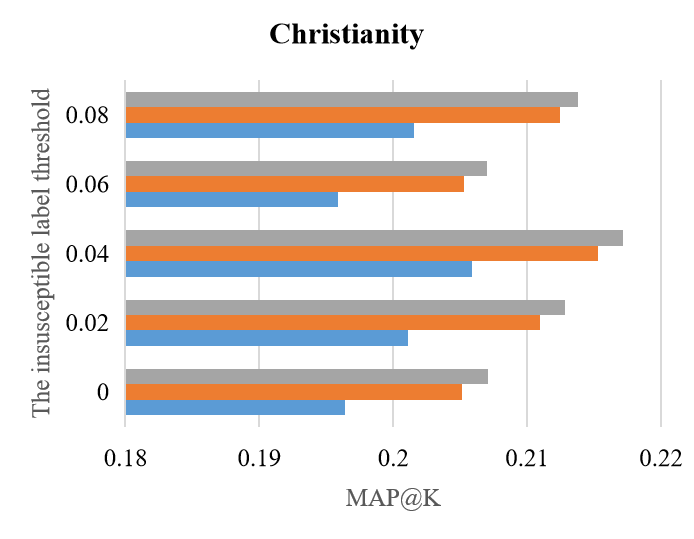}
\end{minipage}
}
\caption{Results of the impact of the insusceptible label threshold on the four public datasets.}   

\label{fig:fig7}    
\end{figure*}

\subsection{Parameter Analysis}
In this part, we further conduct some sensitivity analysis experiments of key parameters on the four datasets to identify how they influence the prediction performance. 

\subsubsection{Impact of the historical cascades proportion}
Since the global cascading graph is constructed based on the historical cascades, which is confirmed as an effective component. Here we further explore how they affect the prediction performance when the historical cascades proportion changes across the range of \{0, 0.2, 0.4, 0.6, 0.8, 1\}, where 0 denotes that no historical cascades are given, while 1 denotes that all the cascades of the training set are given to construct the graph.



The results are reported in Fig.~\ref{fig:fig6}. We observe that as the proportion increases, the prediction performance of the model improves. 
More specifically, for the social media-based datasets i.e. Twitter and Douban, there is a significant increase in performance between 0 and 0.2, and the trend gradually stabilizes afterward.
For the Android dataset, the overall performance improvement is relatively slow. A similar trend is also reflected in Hits@k of the Christianity dataset, but another metric MAP@k shows a linear increase after 0.4.

\subsubsection{Impact of the number of scales $M$}
In this study, we propose a multi-scale diffusion hypergraphs learning structure and confirm its effectiveness through the ablation study. Here we further discuss how the number of time scales $M$ and the  corresponding numbers of time intervals $\Gamma$ affect the prediction performance. We change the number $M$ across the range of \{1, 3, 5\}. 
For the single scale, we set  the corresponding number of time intervals $\Gamma$ to 8. 
For the three scales we propose in MCDAN, we set the corresponding numbers of time intervals $\Gamma$ to \{4, 8, 16\} as in  Sec.~\ref{overall}.
For the five scales, we set the  corresponding numbers of time intervals $\Gamma$ to \{2, 4, 8, 16, 32\}.

The results in Table~\ref{tab:hits_pa_num} and Table~\ref{tab:map_pa_num} show that the method with the three scales we apply in the study ($M=3$) achieves the best performance. It is also noted that excessive time scales may increase memory burden and may not necessarily result in optimal performance.

\subsubsection{Impact of the insusceptible label threshold}
Since the insusceptible label threshold may affect the performance, we thus construct experiments to explore the impact. We change the insusceptible label threshold $t_{ratio}$ across the range of  \{0, 0.02, 0.04, 0.06, 0.08\} and record the insusceptible user size of datasets in Table~\ref{tab:set_lr}.

From Fig.~\ref{fig:fig7}, we find that it is not the higher threshold that leads to better performance. For example, for the Douban dataset, compared with 0.08, 0.06 has better performance.  
This is because an excessively large threshold may cause some inactive users to lose access to information, leading to prediction failure.
From the results of different datasets, we also find that the threshold should not be an invariant constant, which needs to be adapted to the dataset.

\section{Conclusion}\label{conclusion}
In this study, we propose a novel Multi-scale Context-enhanced Dynamic Attention Network (MCDAN) for diffusion prediction.  Different from the previous studies which select the infected user unidirectionally, our model aims at predicting the target user by understanding the user's intention. We construct user representations based on global user dependency and context enhancement from three perspectives of user portrait, which are users' global relationships, multi-scale dynamic preference, and user susceptibility. Comprehensive experiments constructed on four public datasets demonstrate the superiority of the proposed model.

\section*{Acknowledgements}
This work is supported by the National Key Research and Development Program of China (2021YFF0901603), and the National Natural Science Foundation of China (U21B2024, 62202329).

\balance
\bibliographystyle{IEEEtran}	
\bibliography{ref}

\begin{thebibliography}{10}
\providecommand{\url}[1]{#1}
\csname url@samestyle\endcsname
\providecommand{\newblock}{\relax}
\providecommand{\bibinfo}[2]{#2}
\providecommand{\BIBentrySTDinterwordspacing}{\spaceskip=0pt\relax}
\providecommand{\BIBentryALTinterwordstretchfactor}{4}
\providecommand{\BIBentryALTinterwordspacing}{\spaceskip=\fontdimen2\font plus
\BIBentryALTinterwordstretchfactor\fontdimen3\font minus
  \fontdimen4\font\relax}
\providecommand{\BIBforeignlanguage}[2]{{%
\expandafter\ifx\csname l@#1\endcsname\relax
\typeout{** WARNING: IEEEtran.bst: No hyphenation pattern has been}%
\typeout{** loaded for the language `#1'. Using the pattern for}%
\typeout{** the default language instead.}%
\else
\language=\csname l@#1\endcsname
\fi
#2}}
\providecommand{\BIBdecl}{\relax}
\BIBdecl

\bibitem{kumar2022influence}
S.~Kumar, A.~Mallik, A.~Khetarpal, and B.~Panda, ``Influence maximization in
  social networks using graph embedding and graph neural network,''
  \emph{Information Sciences}, vol. 607, pp. 1617--1636, 2022.

\bibitem{oro2017detecting}
E.~Oro, C.~Pizzuti, N.~Procopio, and M.~Ruffolo, ``Detecting topic
  authoritative social media users: a multilayer network approach,'' \emph{IEEE
  Transactions on Multimedia}, vol.~20, no.~5, pp. 1195--1208, 2017.

\bibitem{wang2020social}
H.~Wang, Q.~Meng, J.~Fan, Y.~Li, L.~Cui, X.~Zhao, C.~Peng, G.~Chen, and X.~Du,
  ``Social influence does matter: User action prediction for in-feed
  advertising,'' in \emph{Proceedings of the AAAI conference on artificial
  intelligence}, vol.~34, no.~01, 2020, pp. 246--253.

\bibitem{huang2015social}
S.~Huang, J.~Zhang, L.~Wang, and X.-S. Hua, ``Social friend recommendation
  based on multiple network correlation,'' \emph{IEEE Transactions on
  Multimedia}, vol.~18, no.~2, pp. 287--299, 2015.

\bibitem{zhao2017social}
Z.~Zhao, Q.~Yang, H.~Lu, T.~Weninger, D.~Cai, X.~He, and Y.~Zhuang,
  ``Social-aware movie recommendation via multimodal network learning,''
  \emph{IEEE Transactions on Multimedia}, vol.~20, no.~2, pp. 430--440, 2017.

\bibitem{sang2020context}
L.~Sang, M.~Xu, S.~Qian, M.~Martin, P.~Li, and X.~Wu, ``Context-dependent
  propagating-based video recommendation in multimodal heterogeneous
  information networks,'' \emph{IEEE Transactions on Multimedia}, vol.~23, pp.
  2019--2032, 2020.

\bibitem{feng2022relation}
S.~Feng, C.~Xu, Y.~Zuo, G.~Chen, F.~Lin, and J.~XiaHou, ``Relation-aware
  dynamic attributed graph attention network for stocks recommendation,''
  \emph{Pattern Recognition}, vol. 121, p. 108119, 2022.

\bibitem{zeng2022early}
F.~Zeng and W.~Gao, ``Early rumor detection using neural hawkes process with a
  new benchmark dataset,'' in \emph{Proceedings of the 2022 Conference of the
  North American Chapter of the Association for Computational Linguistics:
  Human Language Technologies}, 2022, pp. 4105--4117.

\bibitem{yang2019multi}
C.~Yang, J.~Tang, M.~Sun, G.~Cui, and Z.~Liu, ``Multi-scale information
  diffusion prediction with reinforced recurrent networks.'' in \emph{IJCAI},
  2019, pp. 4033--4039.

\bibitem{yuan2021dyhgcn}
C.~Yuan, J.~Li, W.~Zhou, Y.~Lu, X.~Zhang, and S.~Hu, ``Dyhgcn: A dynamic
  heterogeneous graph convolutional network to learn users’ dynamic
  preferences for information diffusion prediction,'' in \emph{Machine Learning
  and Knowledge Discovery in Databases: European Conference, ECML PKDD 2020,
  Ghent, Belgium, September 14--18, 2020, Proceedings, Part III}.\hskip 1em
  plus 0.5em minus 0.4em\relax Springer, 2021, pp. 347--363.

\bibitem{sun2022ms}
L.~Sun, Y.~Rao, X.~Zhang, Y.~Lan, and S.~Yu, ``Ms-hgat: Memory-enhanced
  sequential hypergraph attention network for information diffusion
  prediction,'' in \emph{Proceedings of the AAAI Conference on Artificial
  Intelligence}, vol.~36, no.~4, 2022, pp. 4156--4164.

\bibitem{yang2015rain}
Y.~Yang, J.~Tang, C.~Leung, Y.~Sun, Q.~Chen, J.~Li, and Q.~Yang, ``Rain: Social
  role-aware information diffusion,'' in \emph{Proceedings of the AAAI
  Conference on Artificial Intelligence}, vol.~29, no.~1, 2015.

\bibitem{li2013modeling}
D.~Li, Z.~Xu, Y.~Luo, S.~Li, A.~Gupta, K.~Sycara, S.~Luo, L.~Hu, and H.~Chen,
  ``Modeling information diffusion over social networks for temporal dynamic
  prediction,'' in \emph{Proceedings of the 22nd ACM international conference
  on Information \& Knowledge Management}, 2013, pp. 1477--1480.

\bibitem{liao2019popularity}
D.~Liao, J.~Xu, G.~Li, W.~Huang, W.~Liu, and J.~Li, ``Popularity prediction on
  online articles with deep fusion of temporal process and content features,''
  in \emph{Proceedings of the AAAI conference on artificial intelligence},
  vol.~33, no.~01, 2019, pp. 200--207.

\bibitem{gou2018learning}
C.~Gou, H.~Shen, P.~Du, D.~Wu, Y.~Liu, and X.~Cheng, ``Learning sequential
  features for cascade outbreak prediction,'' \emph{Knowledge and Information
  Systems}, vol.~57, pp. 721--739, 2018.

\bibitem{jia2018predicting}
A.~L. Jia, S.~Shen, D.~Li, and S.~Chen, ``Predicting the implicit and the
  explicit video popularity in a user generated content site with enhanced
  social features,'' \emph{Computer Networks}, vol. 140, pp. 112--125, 2018.

\bibitem{yang2019neural}
C.~Yang, M.~Sun, H.~Liu, S.~Han, Z.~Liu, and H.~Luan, ``Neural diffusion model
  for microscopic cascade study,'' \emph{IEEE Transactions on Knowledge and
  Data Engineering}, vol.~33, no.~3, pp. 1128--1139, 2019.

\bibitem{islam2018deepdiffuse}
M.~R. Islam, S.~Muthiah, B.~Adhikari, B.~A. Prakash, and N.~Ramakrishnan,
  ``Deepdiffuse: Predicting the'who'and'when'in cascades,'' in \emph{2018 IEEE
  international conference on data mining (ICDM)}.\hskip 1em plus 0.5em minus
  0.4em\relax IEEE, 2018, pp. 1055--1060.

\bibitem{wang2018sequential}
Z.~Wang, C.~Chen, and W.~Li, ``A sequential neural information diffusion model
  with structure attention,'' in \emph{Proceedings of the 27th ACM
  international conference on information and knowledge management}, 2018, pp.
  1795--1798.

\bibitem{sankar2020inf}
A.~Sankar, X.~Zhang, A.~Krishnan, and J.~Han, ``Inf-vae: A variational
  autoencoder framework to integrate homophily and influence in diffusion
  prediction,'' in \emph{Proceedings of the 13th international conference on
  web search and data mining}, 2020, pp. 510--518.

\bibitem{wang2017linking}
Y.~Wang, X.~Ye, H.~Zhou, H.~Zha, and L.~Song, ``Linking micro event history to
  macro prediction in point process models,'' in \emph{Artificial Intelligence
  and Statistics}.\hskip 1em plus 0.5em minus 0.4em\relax PMLR, 2017, pp.
  1375--1384.

\bibitem{samanta2017lmpp}
B.~Samanta, A.~De, A.~Chakraborty, and N.~Ganguly, ``Lmpp: A large margin point
  process combining reinforcement and competition for modeling hashtag
  popularity.'' in \emph{IJCAI}, 2017, pp. 2679--2685.

\bibitem{zhao2020online}
L.~Zhao, J.~Chen, F.~Chen, F.~Jin, W.~Wang, C.-T. Lu, and N.~Ramakrishnan,
  ``Online flu epidemiological deep modeling on disease contact network,''
  \emph{GeoInformatica}, vol.~24, pp. 443--475, 2020.

\bibitem{lai2020hyfea}
X.~Lai, Y.~Zhang, and W.~Zhang, ``Hyfea: winning solution to social media
  popularity prediction for multimedia grand challenge 2020,'' in
  \emph{Proceedings of the 28th ACM International Conference on Multimedia},
  2020, pp. 4565--4569.

\bibitem{wang2020feature}
K.~Wang, P.~Wang, X.~Chen, Q.~Huang, Z.~Mao, and Y.~Zhang, ``A feature
  generalization framework for social media popularity prediction,'' in
  \emph{Proceedings of the 28th ACM International Conference on Multimedia},
  2020, pp. 4570--4574.

\bibitem{chen2022and}
W.~Chen, C.~Huang, W.~Yuan, X.~Chen, W.~Hu, X.~Zhang, and Y.~Zhang, ``and-tag
  contrastive vision-and-language transformer for social media popularity
  prediction,'' in \emph{Proceedings of the 30th ACM International Conference
  on Multimedia}, 2022, pp. 7008--7012.

\bibitem{wu2022deeply}
J.~Wu, L.~Zhao, D.~Li, C.-W. Xie, S.~Sun, and Y.~Zheng, ``Deeply exploit visual
  and language information for social media popularity prediction,'' in
  \emph{Proceedings of the 30th ACM International Conference on Multimedia},
  2022, pp. 7045--7049.

\bibitem{zhou2021survey}
F.~Zhou, X.~Xu, G.~Trajcevski, and K.~Zhang, ``A survey of information cascade
  analysis: Models, predictions, and recent advances,'' \emph{ACM Computing
  Surveys (CSUR)}, vol.~54, no.~2, pp. 1--36, 2021.

\bibitem{cooper1999inmates}
A.~Cooper, \emph{The inmates are running the asylum}.\hskip 1em plus 0.5em
  minus 0.4em\relax Springer, 1999.

\bibitem{li2022modeling}
G.~Li, W.~Chen, X.~Yan, and L.~Wang, ``Modeling and analysis of group user
  portrait through wechat mini program,'' \emph{Wireless Communications and
  Mobile Computing}, vol. 2022, 2022.

\bibitem{chen2021multi}
Y.~Chen, J.~He, W.~Wei, N.~Zhu, and C.~Yu, ``A multi-model approach for user
  portrait,'' \emph{Future Internet}, vol.~13, no.~6, p. 147, 2021.

\bibitem{zeng2016user}
H.~Zeng and S.~Wu, ``User image and precision marketing on account of big data
  in weibo,'' \emph{Modern Economic Information}, vol.~16, pp. 306--308, 2016.

\bibitem{xing2016user}
L.~Xing, Z.~Song, and Q.~Ma, ``User interest model based on hybrid behaviors
  interest rate,'' \emph{Application Research of Computers}, vol.~33, no.~3,
  pp. 661--665, 2016.

\bibitem{wu2016modeling}
L.~Wu, Y.~Ge, Q.~Liu, E.~Chen, B.~Long, and Z.~Huang, ``Modeling users’
  preferences and social links in social networking services: a joint-evolving
  perspective,'' in \emph{Proceedings of the AAAI conference on artificial
  intelligence}, vol.~30, no.~1, 2016.

\bibitem{gao2016context}
Q.~Gao, D.~Fu, and X.~Dong, ``A context-aware mobile user behavior-based
  neighbor finding approach for preference profile construction,''
  \emph{Sensors}, vol.~16, no.~2, p. 143, 2016.

\bibitem{li2019machine}
J.~Li, S.~Pan, L.~Huang \emph{et~al.}, ``A machine learning based method for
  customer behavior prediction,'' \emph{Tehni{\v{c}}ki vjesnik}, vol.~26,
  no.~6, pp. 1670--1676, 2019.

\bibitem{mueller2016gender}
J.~Mueller and G.~Stumme, ``Gender inference using statistical name
  characteristics in twitter,'' in \emph{Proceedings of the The 3rd
  Multidisciplinary International Social Networks Conference on
  SocialInformatics 2016, Data Science 2016}, 2016, pp. 1--8.

\bibitem{gu2018modeling}
H.~Gu, J.~Wang, Z.~Wang, B.~Zhuang, and F.~Su, ``Modeling of user portrait
  through social media,'' in \emph{2018 IEEE international conference on
  multimedia and expo (ICME)}.\hskip 1em plus 0.5em minus 0.4em\relax IEEE,
  2018, pp. 1--6.

\bibitem{ma2012habit}
H.~Ma, H.~Cao, Q.~Yang, E.~Chen, and J.~Tian, ``A habit mining approach for
  discovering similar mobile users,'' in \emph{Proceedings of the 21st
  international conference on World Wide Web}, 2012, pp. 231--240.

\bibitem{li2010behaviour}
F.~Li, N.~Clarke, M.~Papadaki, and P.~Dowland, ``Behaviour profiling on mobile
  devices,'' in \emph{2010 International conference on emerging security
  technologies}.\hskip 1em plus 0.5em minus 0.4em\relax IEEE, 2010, pp. 77--82.

\bibitem{zhao2016user}
G.~Zhao, X.~Qian, and X.~Xie, ``User-service rating prediction by exploring
  social users' rating behaviors,'' \emph{IEEE Transactions on Multimedia},
  vol.~18, no.~3, pp. 496--506, 2016.

\bibitem{saura2021using}
J.~R. Saura, ``Using data sciences in digital marketing: Framework, methods,
  and performance metrics,'' \emph{Journal of Innovation \& Knowledge}, vol.~6,
  no.~2, pp. 92--102, 2021.

\bibitem{sun2019bert4rec}
F.~Sun, J.~Liu, J.~Wu, C.~Pei, X.~Lin, W.~Ou, and P.~Jiang, ``Bert4rec:
  Sequential recommendation with bidirectional encoder representations from
  transformer,'' in \emph{Proceedings of the 28th ACM international conference
  on information and knowledge management}, 2019, pp. 1441--1450.

\bibitem{kipf2016semi}
T.~N. Kipf and M.~Welling, ``Semi-supervised classification with graph
  convolutional networks,'' \emph{arXiv preprint arXiv:1609.02907}, 2016.

\bibitem{hodas2014simple}
N.~O. Hodas and K.~Lerman, ``The simple rules of social contagion,''
  \emph{Scientific reports}, vol.~4, no.~1, p. 4343, 2014.

\bibitem{zhong2012comsoc}
E.~Zhong, W.~Fan, J.~Wang, L.~Xiao, and Y.~Li, ``Comsoc: adaptive transfer of
  user behaviors over composite social network,'' in \emph{Proceedings of the
  18th ACM SIGKDD international conference on Knowledge discovery and data
  mining}, 2012, pp. 696--704.

\bibitem{Ketkar2021}
\BIBentryALTinterwordspacing
N.~Ketkar and J.~Moolayil, \emph{Introduction to PyTorch}.\hskip 1em plus 0.5em
  minus 0.4em\relax Berkeley, CA: Apress, 2021, pp. 27--91. [Online].
  Available: \url{https://doi.org/10.1007/978-1-4842-5364-9_2}
\BIBentrySTDinterwordspacing

\bibitem{wang2017topological}
J.~Wang, V.~W. Zheng, Z.~Liu, and K.~C.-C. Chang, ``Topological recurrent
  neural network for diffusion prediction,'' in \emph{2017 IEEE international
  conference on data mining (ICDM)}.\hskip 1em plus 0.5em minus 0.4em\relax
  IEEE, 2017, pp. 475--484.

\end{thebibliography}

\end{document}